\documentclass[twocolumn,tighten]{aastex631}
\usepackage{amsmath}

\newcommand{\kms}{km s$^{-1}$}

\usepackage{soul}
\setstcolor{magenta}
\newcommand{\sgra}{Sgr A$^*$ }
\newcommand{\pcc}{ cm$^{-3}$ }
\newcommand{\mpcc}{ $m_{\rm p}$ \pcc}
\newcommand{\ergps}{erg s$^{-1}$ }

\begin{document}
\title{Dissipation of AGN jets in a clumpy interstellar medium}
\shorttitle{Jet-clumpy ISM interactions}

\author{Riju Dutta}\email{rijudutta@iisc.ac.in}
\affiliation{Department of Physics,
Indian Institute of Science,
Bangalore 560012, India}

\author{Prateek Sharma}\email{prateek@iisc.ac.in}
\affiliation{Department of Physics,  
Indian Institute of Science, 
Bangalore 560012, India}

\author{Kartick C. Sarkar}
\affiliation{School of Physics and Astronomy, Tel Aviv University, Tel Aviv, 6997801, Israel}

\affiliation{Dept. of Space, Planetary \& Astronomical Sciences and Engineering, Indian Institute of Technology Kanpur, India}

\author{James M. Stone}
\affiliation{Institute for Advanced Study, 1 Einstein Drive, Princeton, NJ, 08540, USA}

\shortauthors{Dutta et al.}

\begin{abstract}

Accreting supermassive black holes (SMBHs) frequently power jets that interact with the interstellar/circumgalactic medium (ISM/CGM), regulating star-formation in the galaxy. Highly supersonic jets launched by active galactic nuclei (AGN) power a cocoon that confines them and shocks the ambient medium. We build upon the models of narrow conical jets interacting with a smooth ambient medium, to include the effect of dense clouds that are an essential ingredient of a multiphase ISM. The key physical ingredient of this model is that the clouds along the supersonic jet-beam strongly decelerate the jet-head, but the subsonic cocoon easily moves around the clouds without much resistance. We propose scalings for  important physical quantities -- cocoon pressure, head \& cocoon speed, and jet radius. We obtain, for the first time, the analytic condition on clumpiness of the ambient medium for the jet to dissipate within the cocoon and verify it with numerical simulations of conical jets interacting with a uniform ISM with embedded spherical clouds. A jet is defined to be dissipated when the cocoon speed exceeds the speed of the jet-head. We compare our models to more sophisticated numerical simulations, direct observations of jet-ISM interaction (e.g., quasar J1316+1753), and discuss implications for the Fermi/eROSITA bubbles. Our work also motivates effective subgrid models for AGN jet feedback in a clumpy ISM unresolved by the present generation of cosmological galaxy formation simulations.

\end{abstract}


\section{Introduction}

Jets -- which transport energy over large distances in a small solid angle -- are ubiquitous in astrophysics, from non-relativistic protostellar jets (e.g., see \citealt{bally16} for a review) to ultrarelativistic jets launched by stellar (e.g., \citealt{mirabel1998}) and supermassive black holes (for a recent review, see \citealt{blandford2019}). Jet launching appears to be commonly associated with the formation of an accretion disk. After being launched, the AGN jets interact with the ambient medium and this interaction determines the observational appearance of objects such as radio galaxies (see \citealt{saikia2022} for a recent observational review). Radio emission produced by relativistic electrons gyrating around magnetic fields and detected by our radio telescopes has allowed us to put together large catalogs of radio galaxies. These have historically revealed their two broad classes -- the brighter FR-IIs with hot spots at the edge of the radio bubbles and the dimmer FR-Is that are brighter along the path of the jet (\citealt{fanaroff1974}). Recent observations (e.g., \citealt{baldi18, Sabater2019}) show that most massive galaxies have dimmer radio emission at all times.

AGN jets are not only the signposts of supermassive back holes (SMBHs), they also play an active role in regulating star-formation and growth of SMBHs in their host halos. The mechanical energy dumped by these jet in the surrounding ISM and the diffuse intracluster/circumgalactic medium (ICM/CGM) is enough to offset radiative cooling losses of these massive atmospheres (for a review, see \citealt{fabian12}). One of the still unresolved problems in this field is how narrow jets can effectively isotropically heat the ICM/CGM (e.g., see \citealt{Meece2017, Choudhury2022}). To understand the AGN heating mechanism in these systems, we must understand the physics of the interaction of the jet and the ambient medium, the main subject of our paper. 

The theoretical proposals to explain the broad classification of radio galaxies involve both internal and external processes. An example of the former is the magnetic kink instability of the jet, which makes it unstable to helical perturbations and dissipates it along its path, producing an FR-I type appearance (e.g., \citealt{tchekhovskoy2016}). Examples of the other class of models involving the interaction of the jet with a clumpy ISM are  \citet{sutherland2007, mukherjee2016, Mukherjee2018}.

In this paper, we investigate the interaction of a collimated hydrodynamic AGN jet with a clumpy medium with analytic estimates and high-resolution 3D hydrodynamic simulations. In particular, we derive the physical condition on the clumpiness of the ISM for jet dissipation in a clumpy medium. Jet is dissipated when the jet-head is confined within a roughly isotropic cocoon. Our scaling for jet dissipation can be applied in various contexts, from a proper implementation of jet feedback in cosmological galaxy formation simulations to explaining the appearance of radio galaxies.

An analytical model for jet-head propagation and cocoon expansion as a function of time in a smooth medium was given by \cite{bromberg11}. Their main conclusion is that collimated jets propagating in a uniform medium always give rise to narrow cocoons. This conclusion, however, would need to be modified in a clumpy medium with a large density contrast between the clouds and the low-density diffuse phase. In such a medium, where the jet is obstructed by dense clouds for a large fraction of the propagation time, the jet-head encounters a relatively high effective density. On the other hand, the over-pressured cocoon whose internal velocities are much smaller than its sound speed expands faster through the low-density diffuse phase that fills the space between the clouds. So it encounters a much lower effective density.\footnote{The qualitative difference between clouds interacting with the jet and the clouds interacting with the cocoon is shown in the following video of an equivalent shallow water setup: \url{http://youtu.be/DUpSwMMrGfk}.} This effect leads to cocoons that are much wider and isotropic than expected in a uniform medium. Therefore, the calculations of \cite{bromberg11} would need to be modified in two ways. First, instead of the ambient density \(\rho_H\), we need to have two different effective ambient densities for head propagation and cocoon expansion. Second, the volume of the cocoon gas would need to be multiplied by a factor of \(1-f_V\), where \(f_V\) is the volume filling factor of the dense clouds inside the cocoon. 

The above considerations allow us to formulate, for the first time, a criterion for a jet to be dissipated in a clumpy medium. We present analytical estimates of the dissipation criteria for a simple two-phase medium where dense clouds are distributed uniformly. We find that the dissipation criterion only depends on the volume filling factor and the density contrast of the clouds. We verify our analytical estimates using high-resolution numerical simulations and find that the analytical criterion for dissipation works well except that we need to calibrate the size of the cocoon with numerical simulations. Our dissipation criterion, therefore, can be used in a wide range of systems to understand jet-ISM interaction and even provide predictions for the evolution of the system.

In section \ref{sec:setup} we present the physical and numerical setup of our jet-ISM interaction model. Section \ref{sec:analytic_estimates} presents analytic estimates for key physical quantities (e.g., cocoon pressure, head/cocoon speed) in different jet-ISM interaction scenarios and the jet dissipation criterion. Section \ref{sec:results} tests the analytic models with appropriately designed hydrodynamic simulations. Section \ref{sec:astro_implications} discusses the broad astrophysical implications of our work, and section \ref{sec:summary} is a summary of our paper.

\section{Setup}
\label{sec:setup}
The ISM around the AGN jets can be very complex in geometry and temperature/density distribution (see  \citealt{Heywood2022} \& references therein). Therefore, we consider the simplest model of a clumpy medium -- namely a uniform medium with homogeneous density in which overdense spherical clouds of radius $R_{\rm cl}$ are spread randomly in pressure equilibrium with the intracloud medium. This oversimplification makes the setup somewhat analytically tractable. Insights gained from this simple model can be extended to a more realistic situation.

\subsection{Physical setup}
\label{sec:physical_setup}
First, we generalize the analytic model of \citet{bromberg11} for the structure of the cocoon driven by narrow AGN jets in a uniform medium to the non-relativistic jet regime (section \ref{sec:uniform_analytic}), and calibrate it with numerical simulations (section \ref{sec:uniform_test}). Next, we treat the impact of clouds on jet and cocoon evolution both analytically (section \ref{sec:jet_clumpy_analytic}) and with numerical simulations (section \ref{sec:jets_clumpy}). 

The clumpy medium that we consider consists of a volume-filling warm phase with a constant density $\rho_H$ = $1\ m_p$ cm$^{-3}$ and temperature $10^4$ K (mimicking the warm/ionized diffuse phase of the ISM), within which uniformly distributed spherical dense clouds are positioned randomly. The density and volume filling fraction of the clouds are denoted by $\rho_{\rm cl}$ and $f_V$, respectively. The clouds are in pressure equilibrium with the diffuse warm phase and have a fiducial radius of $5$ pc.

We inject a conical jet with a half-opening angle ($\theta_0$) of $10^{\circ}$ into this medium. The radius of the jet at its injection height is 2 pc, which is small compared to the collimated radius of the jet $R_j$ for our runs. The total power of the jet ($L_j$) is $1.1\times10^{42}$ erg s$^{-1}$, which is dominated by its kinetic energy (the thermal power being only $10\%$ of the kinetic power at the point of injection). The initial jet velocity ($v_j$) is assumed to be $1.5\times10^4$ km s$^{-1}$ ($c/20$, where $c$ is the speed of light)
which is highly supersonic compared to the ambient medium with a sound speed $\approx 10$ km s$^{-1}$. 
Table \ref{table:jet_sim_parameters} lists our parameter values for a quick reference. 

\begin{table} 
\begin{center}
\scalebox{0.87}{
\hspace*{-1.7cm}\begin{tabular}{c c c c c c}
\hline
$\theta_0$ & $L_j$  & $v_j$ & $\rho_H$ & $T_H$& $L_{\rm box}$ \\
  & (erg s$^{-1}$) &  & ($m_p$ cm$^{-3}$) & (K) & (kpc)  \\
\hline \hline
$10^\circ$ & 1.1$\times 10^{42}$ & 0.05 $c$ & 1 & $10^4$ &  0.8 \\
  &   &   &   &   &  0.4 \\
\hline
\end{tabular}
}
\end{center}
\caption{\label{table:jet_sim_parameters} The jet \& homogeneous ISM parameters used in most simulations. Cubical boxes with sides 0.4 and 0.8 kpc are used for different simulations listed in Tables \ref{table:rho_const} \& \ref{table:pattern_const}).}
\end{table}

For this work, the parameters governing the jet injection ($\theta_0$, $L_j$, and $v_j$) are fixed at the above-mentioned values for all runs. We vary the parameters describing the ambient medium -- the cloud size ($R_{\rm cl}$),  cloud density ($\rho_{\rm cl}$), and volume fraction of clouds ($f_V$). However, our analytic scaling relations allow us to apply our findings to a general choice of parameters.

\subsection{Numerical setup}
\label{sec:numerical_setup}
For our hydrodynamic simulations, we use the AthenaK code (\citealt{Stone2020}; Stone et al. in prep.) with a Harten-Lax-van Leer-Einfeldt (HLLE) Riemann solver, RK2 time integration and piecewise linear (PLM) reconstruction. The code solves the following fluid equations in their conservative form.
\begin{subequations}
	\begin{align}
	\label{eq:continuity}
	&\frac{\partial\rho}{\partial t}+\nabla\cdot (\rho \bm{v})=0,\\
	\label{eq:momentum}
	&\frac{\partial(\rho\bm{v})}{\partial t}+\nabla\cdot (\rho\bm{v}\otimes\bm{v})+ \nabla P= 0,\\
	\label{eq:energy}
	&\frac{\partial E}{\partial t}+\nabla\cdot ((E+P)\bm{v})= 0,\\
	\label{eq:tot_energy}
	&E=\frac{\rho\bm{v}\cdot\bm{v}}{2} + \frac{P}{\gamma-1}.
	\end{align}
\end{subequations}
Here \(\rho\) is the mass density, \(P\) is the thermal pressure and \(\bm{v}\) is the velocity of the gas at any given point in space and time, and \(\gamma\) is the adiabatic index of the ideal gas. We do not include gravity, since it affects the dynamics on larger time/length scales than studied here. We also ignore radiative cooling for the current work to focus on the hydrodynamic interactions. We discuss the effect of neglecting radiative cooling in section \ref{subsec:excluded-physics}.

\subsubsection{Grid \& boundary conditions}
\label{sec:grid}
The simulation domain is taken to be a cubical box of side $L_{\rm box}= 0.8$ and $0.4$ kpc for two sets of runs listed in Tables \ref{table:rho_const} and \ref{table:pattern_const}, respectively. The outer regions have a base resolution of 512 cells along each side (i.e., 1.28 cells/pc for $L_{\rm box} = 0.4$ kpc, and half this value for the larger boxes). In the inner regions, close to the axis of the jet (z-axis), we employ static mesh refinement (SMR). Due to the SMR, the region defined by $|x| < 25$ pc and $|y| < 25$ pc ($|x| < 50$ pc and $|y| < 50$ pc for $L_{\rm box} = 0.8$ kpc) has a $4\times$ higher resolution (of 5.12 cells/pc for $L_{\rm box} = 0.4$ kpc and 2.56 cells/pc for $L_{\rm box} = 0.8$ kpc). Thus, the clouds in the path of the jet have nearly 50 (25) cells across their diameter for $R_{\rm cl}=5$ pc and $L_{\rm box} = 0.4$ ($0.8$) kpc. This ensures that the jet collimation shock and jet-cloud interactions are sufficiently resolved. We run each simulation till the cocoon or the jet head moves out of the computational domain. For the larger box simulations, with $L_{\rm box} = 0.8$ kpc, the same mesh pattern is used such that the resolution everywhere is lowered by a factor of 2.

The jet is injected at the bottom boundary with a velocity at an angle $\theta_0$ relative to the $+z$-direction, within a jet injection radius of 2 pc (same for both $L_{\rm box} = 0.8$ and 0.4 runs).\footnote{This means that the jet injection region is resolved only by $\approx$ 5 grid cell for $L_{\rm box} = 0.8$ kpc and the injected jet power can be off from the value mentioned in Table \ref{table:jet_sim_parameters}. This is tolerable since the jet dissipation criterion depends only upon the jet angle $\theta_0$ and the clumpiness (c.f. Eq. \ref{eq:calibrated_lambda_crit}), and not on jet power.} At the lower z-boundary corresponding to the base of the jet, we impose outflow boundary conditions (zero gradient) inside the injection radius of 2 pc, and reflective boundary conditions outside it. At all the other boundaries, we impose outflow boundary conditions. The reflective boundary condition at the bottom boundary outside the jet injection region mimics the interaction of the cocoon with its counterpart in the $-z$ domain. We run our simulations until the cocoon outer shock reaches the lateral edges of the box or the jet head reaches the upper z-boundary. We restrict all our measurements within this time.

\begin{figure*}
\begin{centering}
{\mbox{\includegraphics[width=\textwidth]{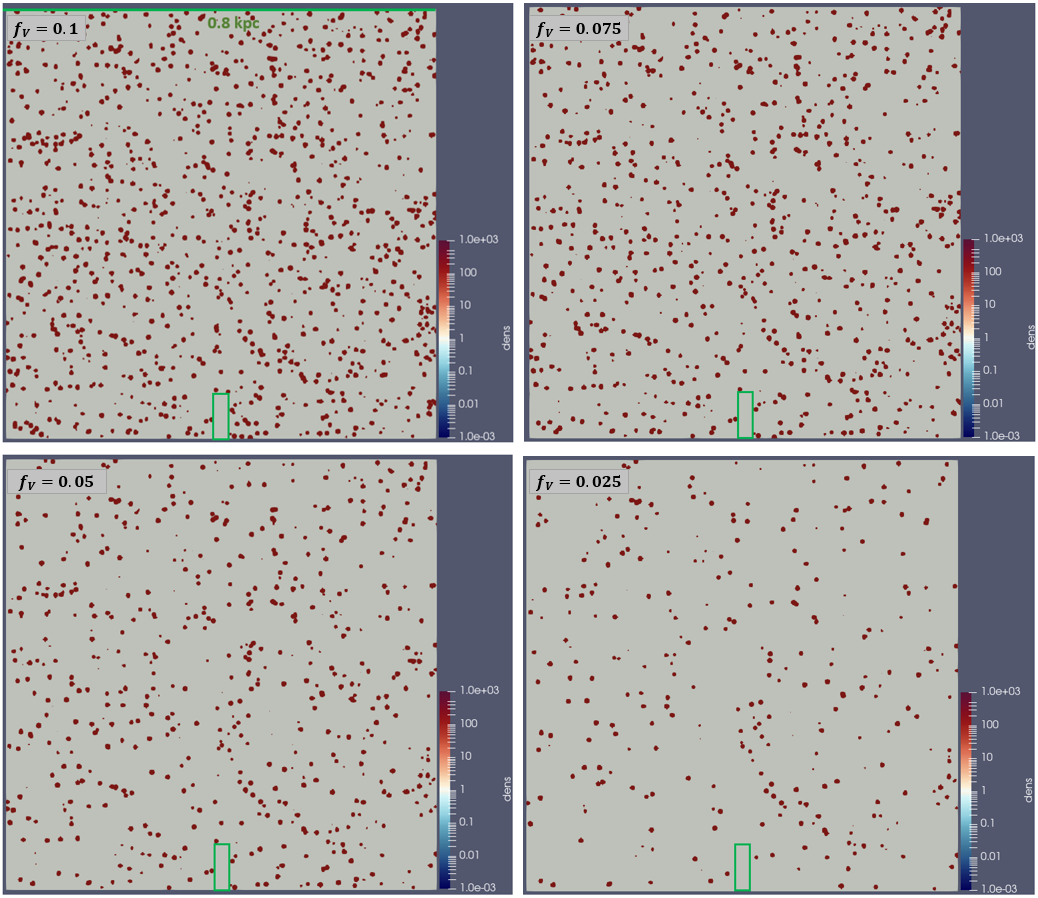}}}
\end{centering}
\caption{Snapshots of density (in units of \mpcc) in y-z plane for $L_{\rm box} = 0.8$ kpc simulations with the same cloud pattern (Table \ref{table:rho_const}) but a varying volume fraction of $f_V=0.1,~0.075,~0.05$ and $0.025$. The cloud density $\rho_{\rm cl}$ is 100 $m_p\ {\rm cm}^{-3}$ and cloud-size $R_{\rm cl}=5 $ pc. For comparison with analytic scalings, we maintain the same cloud locations but remove some clouds across these runs (see section \ref{sec:cloud_distribution}). Also note that for faster collimation of the conical jet, we hollow out a small cuboidal volume region close to the center (shown in green) that does not contain any clouds.
} 
\label{fig:initial_hollowed}
\end{figure*}

\subsubsection{Cloud distribution}
\label{sec:cloud_distribution}
To carefully compare our simulations with analytic models, we run two kinds of simulations: ones with a fixed density but a changing cloud pattern (i.e. changing the volume filling fraction and cloud radius, as listed in Table \ref{table:rho_const}), and ones with a changing density for a fixed cloud pattern (listed in Table \ref{table:pattern_const}). For the former, we hollow out a small cuboidal volume close to the center, of horizontal width $14\ {\rm pc}$ and height $85\ {\rm pc}$, where we do not place any clouds (see small green rectangles in panels of Fig. \ref{fig:initial_hollowed}). This allows the collimation shock to form without interruption, which gives us enough time to measure the cocoon and head properties before they leave the box. Such hollowing out does not affect the later evolution of jets after the formation of the collimation shock but allows us to unambiguously compare simulations with analytic scalings. The runs with changing cloud density do not impose this hollowing out. 

The specific location of clouds can affect the jet head and cocoon properties at certain times. To avoid statistical fluctuations and to measure trends with cloud properties, we place the clouds at identical locations and remove a specified random fraction of them as we decrease the volume fraction or increase the cloud radius for a fixed volume fraction ($f_V$; see Fig. \ref{fig:initial_hollowed} for an example). Cloud locations are the same for the runs within Tables \ref{table:rho_const} and \ref{table:pattern_const}, but not across them.
%{\bf PS: refer to some figures to show this clearly.}

\subsubsection{Measuring jet-head \& cocoon positions} 
\label{sec:measure_head_cocoon}
At time steps separated by 0.02 Myr, we analyze 2-D slices of the box in the y-z plane ($x = 0$). From these slices, we extract the lateral size of the cocoon ($r_c$) and the height of the jet-head ($z_h$). We calculate $r_c$ as the average of the maximum lateral extent of the cocoon in the +y and -y directions (the two values are quite similar in a medium where the cloud distribution is statistically homogeneous and isotropic on the scale of the cocoon). We measure $z_h$ to be the length of the high velocity ($v_z > 10^4$ \kms) jet. In some cases, where the jet is obstructed by a cloud and there is no clear jet head, we assume the cloud location to be the $z_h$. The scales $r_c$ and $z_h$ are measured at intermediate times after the jet becomes collimated but before any of the cocoon material leaves the box. 

\section{Analytic Estimates}
\label{sec:analytic_estimates}

Before moving on to 3D hydrodynamic simulations of jet-ISM interaction, we provide some analytic estimates. We start with a recapitulation of the analytic model of jet collimation in a homogeneous medium from \citet{bromberg11} and generalize it to nonrelativistic jets (section \ref{sec:uniform_analytic}). In section \ref{sec:jet_clumpy_analytic} we generalize these estimates to a clumpy medium.

\subsection{Collimated jet in a uniform medium}
\label{sec:uniform_analytic}
Even the earliest works on AGN jets recognized the anisotropy of an AGN jet expanding into a uniform medium (e.g., \citealt{blandford1974,scheuer1974}). Later works (e.g., \citealt{begelman1989,matzner2003}) calculated the velocity of the head pushed by the jet ram pressure and the cocoon velocity driven by the cocoon pressure but the width of the collimated jet was not calculated self consistently. Building on the earlier works on the structure of jet collimation shock (e.g., \citealt{komissarov1997}), \citet{bromberg11} presented a model for the structure of the cocoon blown by jets interacting with a smooth medium. 

We assume a steady cold jet with power \(L_j\), jet velocity $v_j$, and the jet injection angle \(\theta_0\). These three parameters define all the jet properties. The cold ambient medium is assumed to have a uniform density $\rho_H$.\footnote{Our results, for both smooth and clumpy ISM, can be easily generalized to a power-law profile of the ambient medium following \citet{bromberg11}.} The jet-head speed is $v_h$, the cocoon expansion speed \(v_c\), and the cocoon pressure \(P_c\). {Although the} calculations presented here are for a non-relativistic jet, the scaling relations derived should apply to mildly relativistic jets, and qualitatively to even relativistic jets.

{For a collimated jet propagation into an ambient medium, the jet terminates (or creates a reverse shock) just behind the forward shock (see figure 3 of \cite{Sarkar2023} for a qualitative depiction). Therefore, we can apply momentum flux conservation along the jet direction, which in the forward shock frame is}
\begin{equation}
\rho_j (v_j - v_h)^2 + P_j = \rho_H {v_h}^2 + P_a,
\end{equation}
where \(P_j\) and \(P_a\) are the pressures in the jet and ambient medium respectively, and $\rho_j$ is the jet density just upstream of the head. The jet density, $\rho_j$, and velocity, $v_j$, are roughly constant between the collimation shock and the jet-head since the jet quickly becomes cylindrical after the collimation shock. Now assuming a strong forward shock and kinetic energy dominated jet, the pressures can be neglected, leading to
\begin{equation} \label{eq:v_h}
v_h = \frac{v_j}{1 + \sqrt{\frac{\rho_H}{\rho_j}}} \approx \sqrt{\frac{\rho_j}{\rho_H}} v_j,
\end{equation}
where in the last equality, we assume that \(\rho_j \ll \rho_H\). Such an assumption is appropriate for the AGN jets which are generally of low density. One needs to relax this assumption in cases where the jets may have higher density, such as proto-stellar jets.

For the cocoon expansion (perpendicular to the jet direction), we again consider momentum flux conservation in the frame of the horizontal forward shock. Assuming a strong shock, we get
\begin{equation}
\frac{1}{4} \rho_H {v_c}^2 + P_c = \rho_H {v_c}^2,
\end{equation}
so that,
\begin{equation} 
\label{eq:v_c}
v_c = \sqrt{\frac{4}{3} \frac{P_c}{\rho_H}}.
\end{equation}

In the second step, using these expressions obtained for \(v_h\) and \(v_c\) and assuming a cylindrical geometry for the entire cocoon, we get the following estimate for the volume of the cocoon
\begin{equation} \label{eq:volume_uniform}
V_c \approx \pi v_c^2 v_h t^3
\end{equation}
where \(t\) is time. {Therefore,} the pressure within the cocoon, $P_c \approx (2/3)L_j t/V_c$. Now, combining with Eq. \ref{eq:v_c}, we obtain the expression for the cocoon pressure to be
\begin{equation}
\label{eq:P_c}
    P_c = \sqrt{\frac{L_j \rho_H}{2 \pi v_h t^2}}.
\end{equation}

The key advance in \citet{bromberg11} was to include the expression for the jet radius, $R_j$, by solving for the structure of the jet collimation shock. {It is found that the collimation shock turns an initially conical jet into a cylindrical one after a certain height, $H_t/2$, } and the jet speed and the jet density are roughly constant along the cylindrical jet-beam. Following \citet{bromberg11} and generalizing their calculation to a non-relativistic jet (see their Eq. 9), the height of the tip of the collimation shock is given as 
\begin{equation}
\label{eq:Ht}
H_t = \sqrt{\frac{2 L_j}{\pi v_j P_c}},
\end{equation}
such that the jet radius is $R_j = \theta_0 H_t/2$. Therefore, the cross-section area of the jet-head, $\Sigma_j$, is given by
\begin{equation} 
\label{eq: Sigma_j}
\Sigma_j \equiv \pi R_j^2 = \frac{L_j \theta_0^2}{2 P_c v_j}.
\end{equation}
The jet pressure outside the collimation shock equals the cocoon pressure. The jet density in the collimated cylindrical jet (with a roughly constant radius $R_j$) equals (using Eq. \ref{eq: Sigma_j} and expressing jet luminosity in terms of its density; $L_j = 0.5 \rho_j v_j^3 \pi R_j^2$)
\begin{equation}
\label{eq:rho_j}
  \rho_j \approx \frac{4 P_c}{v_j^2 \theta_0^2}.  
\end{equation}
Combining Eqs. \ref{eq:rho_j}, \ref{eq:P_c} \& \ref{eq:v_h}, we obtain
\begin{subequations}
\label{eq:analytical_pred_all}
\begin{align}
    & v_h \approx 2  \left( \frac{L_j}{4\pi \theta_0^4 \rho_H t^2} \right)^{1/5} \nonumber\\
    & \ \ \ \approx 2.8 \times 10^3~{\rm km~s^{-1}}  \left(\frac{L_{j,42}}{\theta_{0,10}^4 \rho_{H, m_p} t_{\rm Myr}^2}\right)^{1/5}
    \label{eq:vh_uniform_pred}\\
   & v_c \approx \sqrt{\frac{4}{3}} \left( \frac{\theta_0 L_j}{4 \pi \rho_H t^2} \right)^{1/5} \nonumber\\
   & \ \ \ \approx 2.8 \times 10^2~{\rm km~s^{-1}} \left(\frac{\theta_{0,10} L_{j,42}}{\rho_{H, m_p} t_{\rm Myr}^2}\right)^{1/5}
   \label{eq:vc_uniform_pred}\\
   &P_c \approx  \left( \frac{\theta_0^2 L_j^2 \rho_H^3}{16\pi^2 t^4} \right)^{1/5} \nonumber\\
   & \ \ \ \approx 9.8 \times 10^{-10}~{\rm dyn~cm^{-2}} \left(\frac{\theta_{0,10}^2 L_{j,42}^2\rho_{H, m_p}^3}{t_{\rm Myr}^4}\right)^{1/5} 
   \label{eq:P_uniform_pred}
\end{align}
\end{subequations}
where {$L_{j,42} = L_j/(10^{42}$ \ergps), $\theta_{0,10} = \theta_0/10^\circ$, $\rho_{H,m_p} = \rho_H/m_p$, and $t_{\rm Myr} = t$/Myr}. Note that the cocoon and head properties only depend on the usual combination of $L_j$, $\rho_H$ and $t$ as in the standard wind-bubble solution (\citealt{Weaver1977}) but with an additional dependence on $\theta_0$ such that 
\begin{equation}
\label{eq:vh-vc-uniform-medium}
    \frac{v_c}{v_h} \approx \frac{\theta_{0,10}}{10} \,.
\end{equation}
Now from Eq.  \ref{eq: Sigma_j}, we find that the collimated jet radius evolves as
\begin{equation}
\begin{split}
\label{eq:Rj_uniform_pred}
&R_j \approx  \left( \frac{\theta_0^8 L_j^3 t^4}{2 \pi^3 v_j^5\rho^3_a}\right)^{1/10}  \\
&\ \ \ \approx 4.16~{\rm pc} \left( \frac{v_j}{c}\right)^{-1/2} \left( \frac{\theta_{0,10}^8  L_{j,42}^3 t_{\rm Myr}^4}{{\rho^3_{a, m_p}}}\right)^{1/10},
\end{split}
\end{equation}
which depends also on the jet velocity unlike the head/cocoon parameters (Eqs. \ref{eq:vh_uniform_pred}-\ref{eq:P_uniform_pred}). Eq. \ref{eq:Rj_uniform_pred} shows that even for a jet speed $v_j \approx 0.01c$, the self-consistently collimated jet radius is $\sim 10$s of pc. So resolving the collimation shock is not possible for large-scale galaxy formation simulations, hence the jets must be included as a subgrid model (see section \ref{sec:jets_cosmo}).

The analytic estimates above show that the jet+cocoon evolution is qualitatively different from the evolution of a spherical wind-blown bubble. As with the wind, the cocoon properties ($v_h$, $v_c$, $P_c$) depend on the appropriate dimensional combinations of $L_j$, $\rho_H$, $t$, but there is also a dependence on another dimensionless parameter, the jet opening angle $\theta_0$. The dependence on $\theta_0$ cannot come from dimensional arguments but from physical considerations. The height of the collimation shock ($H_t$) and jet radius ($R_j$; see Eq. \ref{eq:Rj_uniform_pred})  depend on the jet velocity, similar to the termination shock radius for a spherical wind which also depends on the ejecta velocity (e.g., Eq. 9 in \citealt{Sharma2014}). 

\begin{figure*}
\begin{centering}
\hspace*{-0.9cm}\begin{tabular}{c c c}
{\mbox{\includegraphics[width=0.33\textwidth]{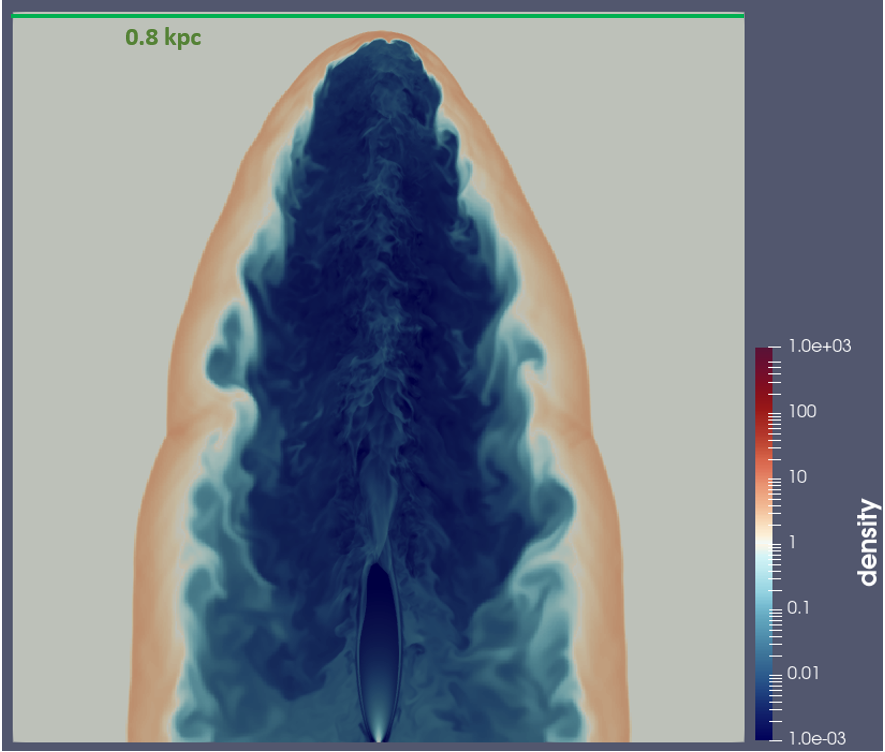}}}
{\mbox{\includegraphics[width=0.33\textwidth]{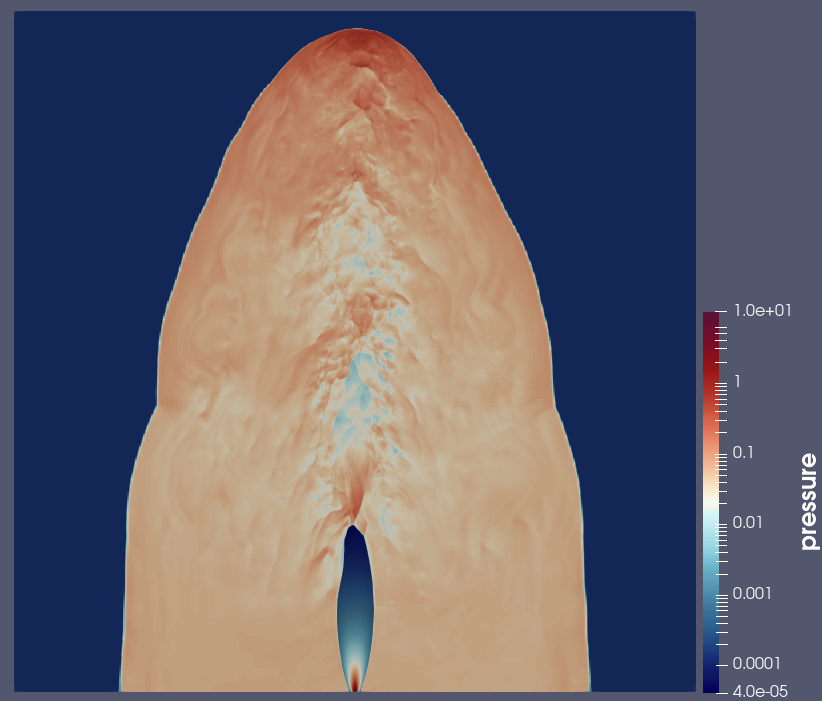}}}
{\mbox{\includegraphics[width=0.33\textwidth]{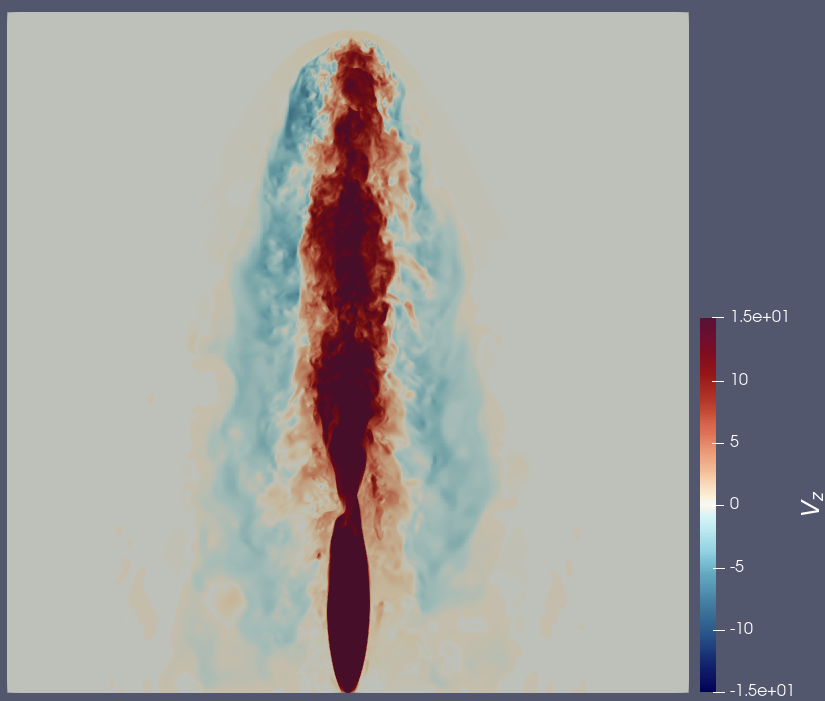}}}\\
\end{tabular}
\end{centering}
\caption{Snapshots in y-z plane of density ($m_p$ cm$^{-3}$, left), pressure ($1.67 \times 10^{-8}\ {\rm dyn~cm}^{-2}$, center) and vertical velocity (1000 km s$^{-1}$, right) for jet in a uniform medium with $\rho_H = 1\ m_p$ cm$^{-3}$, at time $t = 0.6$ Myr. Note the presence of the converging collimation shock driven by the overpressured cocoon, as described in \citet{bromberg11}. This brings the jet into pressure equilibrium with the cocoon downstream of the convergence point. The shocked ISM is compressed in a dense shell which shows signs of shear instabilities at the inner boundary, pressure is almost constant throughout the cocoon, and there are backflows in the inner cocoon deflected from the jet-head.
} 
\label{fig:uniform_end}
\end{figure*}

One of the key predictions of this jet-cocoon model is that for a collimated jet, the aspect ratio of the cocoon, which can be approximated by $v_c/v_h$, is of the order of $\theta_0$. This implies that such jets always produce narrow cocoons, and hence are not dissipated in a smooth medium. Figure \ref{fig:uniform_end} shows the jet and cocoon structure for a jet expanding into a uniform medium.

\subsection{Interaction of a jet/cocoon with single/few clouds}
Before studying the interaction of a jet with numerous clouds, it is useful to estimate some timescales for the interaction of a jet with a single cloud. The problem of the interaction of a supersonic uniform wind with a dense/cold cloud (the cloud-crushing problem), which applies to the jet-cloud interaction problem for $R_{\rm cl} \ll R_j$, has been studied exhaustively (\citealt{Klein1994}) but quantitative understanding is still lacking (\citealt{Ranjan2011}). The key conclusion from non-radiative simulations is that the cloud is shredded by compression and hydrodynamic (Kelvin-Helmholtz, Rayleigh-Taylor, and Richtmyer-Meshkov) instabilities over a few cloud-crushing times ($t_{\rm cc} = \sqrt{\rho_{\rm cl}/\rho_w} R_{\rm cl}/v_w$, where $\rho_w$ is the wind density and $v_w$ is the wind speed), much faster than it can become comoving over a drag timescale ($t_{\rm drag} = [\rho_{\rm cl}/\rho_w] R_{\rm cl}/v_w$).\footnote{However, see \citealt{Goldsmith2018, Forbes2019} who find the momentum exchange time $\sim t_{\rm cc}$. This is understandable since the density contrast disappears after the cloud is mixed, and the cloud becomes comoving immediately after this.}

A smaller number of works, mostly in the context of dense protostellar jets, have studied the interaction of jets narrower than or comparable to the cloud size (e.g., \citealt{Raga1996,dGDP1999,Wang2000}). These works show that the off-centered jets are deflected for some time before they clear out the cloud material from their path. For $R_j < R_{\rm cl}$, a jet impacting close to the center launches a shock into the cloud and the jet-head moves slowly through the cloud at a head velocity $v_h \approx \sqrt{\rho_j/\rho_{\rm cl}} v_j$ (analogous to Eq. \ref{eq:v_h}). Thus, after $\approx 2 R_{\rm cl}/v_h = 2 t_{\rm cc}$ the jet is expected to drill through the cloud, leaving behind the cloud material which is pushed laterally by the cocoon. In this case too, the drag time is $t_{\rm drag} =  (\rho_{\rm cl}/\rho_w) R_{\rm cl}/v_w$, longer than $t_{\rm cc}$. But in contrast to a cloud smaller than the jet-beam, in this case, the jet-head can drill through the cloud, leaving behind most of the cloud mass that is not comoving with the jet head.

With multiple clouds interacting with the jet, collective effects can become important, and the cloud destruction can be significantly delayed (e.g., \citealt{Forbes2019}). Similarly, the destruction timescale can have a non-trivial dependence on $R_{\rm cl}/R_j$ for cloud size comparable to the jet radius. Irrespective of these complications which are difficult to treat from first principles, in all cases the cloud material along the jet-beam interacts strongly with the jet, but the cocoon essentially moves around the dense clouds. The clouds facing the jet-beam are ablated and/or moved away from its path. In contrast, the clouds engulfed by the cocoon only encounter the shock for a short time ($\sim R_{\rm cl}/v_c$), after which they are embedded in the subsonic, low-density cocoon which moves through the diffuse phase avoiding the clouds. These concepts from jet-single/few cloud interaction motivate the timescale estimate presented in section \ref{sec:uniform_clouds_analytic} for the interaction of a jet with uniformly distributed clouds.

\subsection{Collimated jet in a clumpy medium}
\label{sec:jet_clumpy_analytic}
{I}t is useful to generalize the results from section \ref{sec:uniform_analytic} to a more realistic setting of a clumpy ISM. A cloud present in the jet beam encounters a supersonic flow (jet velocity $\gg$ jet sound speed), leading to the formation of a strong bow shock with a bow wave  angle $\sim \sin^{-1}(1/M_j)$, where $M_j = v_j/c_{s,j}$ is the jet Mach number. The bow shock persists till the cloud is fully ablated. 
{On the other hand, the clouds inside the cocoon only face a subsonic flow and therefore do not create bow shocks. The subsonic cocoon material simply flows around the dense cloud.} 
Thus, in a clumpy ISM with a small volume filling fraction of clouds, the supersonic jet 
would on average encounter material with an effective density, $\rho_{\rm eff}$, that is higher than the density of the homogeneous warm phase, $\rho_H$. Meanwhile, the over-pressured subsonic cocoon can expand through the warm phase much faster, such that the relevant ambient density for cocoon expansion is $\rho_H$ (similar to the propagation of a supernova blast wave in a clumpy ISM; e.g., \citealt{McKee1977}). This leads to a head velocity that is lower than in a uniform medium of density $\rho_H$, leading to a smaller $v_h/v_c$. {Now, if there are too many dense clouds along the jet path, the average $v_h/v_c$ will be smaller. At a critical number density of the clouds (or a critical value of $\rho_{\rm eff}$), the ratio can even become less than one. After this point, the pressure-driven cocoon overtakes the ram pressure-driven jet head even in the direction of the jet, and the whole cocoon expands in a near-spherical fashion. The jet is then said to be dissipated, since the dynamics of the cocoon is no longer driven by the kinetic energy of the jet.}

To model jet interaction with a clumpy medium, we begin by defining {an effective} density contrast, $\lambda$ as 
\begin{equation}
\label{eq:lambda_def}
\lambda \equiv \frac{\rho_{\rm eff}}{\rho_H},
\end{equation}
where $\rho_{\rm eff}$ (to be specified later), the effective density encountered by the jet-head{. As we will see later, this effective density} is a function of the cloud size, volume filling fraction of the clouds ($f_V$), and the cloud density $\rho_{\rm cl}$. The new expressions for $v_h$ and $v_c$ (that replace Eqs. \ref{eq:v_h} \& \ref{eq:v_c}) would therefore be
\begin{equation} \label{eq:new_v_h}
v_h = \sqrt{\frac{\rho_j}{\rho_{\rm eff}}} v_j, 
\end{equation}
\begin{equation} \label{eq:new_v_c}
v_c = \sqrt{\frac{4}{3} \frac{P_c}{\rho_H}}.
\end{equation}
Now in this model, Eq. \ref{eq:volume_uniform} for the volume of the cocoon would also need to be modified to exclude the volume occupied by clouds, and the resulting expressions would be different for the non-dissipated (cylindrical cocoon) and dissipated (spherical cocoon) regimes.

\subsubsection{Non-dissipated jet}
\label{sec:analytic_nondiss}
In the non-dissipated case, where $v_h > v_c$, the vertical expansion of the cocoon is still determined by the head velocity as in the uniform case. So the volume occupied by the cocoon gas is
\begin{equation}
V_c \approx (1-f_V) \pi v_c^2 v_h t^3,
\end{equation}
which is just the volume enclosed by the outer cocoon minus the volume occupied by clouds inside the cocoon. This assumes that the clouds are destroyed by the cocoon 
on a timescale longer than the relevant dynamical timescale (since flow inside the cocoon {is} subsonic), so that the clouds remain intact/static for much longer. Moreover, the expression for jet density in terms of cocoon pressure remains the same as Eq. \ref{eq:rho_j}. With these modifications, the expressions for the time evolution of the head and cocoon parameters become,
\begin{subequations}
\begin{align}
&v_h \approx 2 (4 \pi)^{-1/5} (1-f_V)^{-1/5} \lambda^{-2/5}  \left( \frac{L_j}{\theta_0^4 \rho_H t^2} \right)^{1/5} \label{eq:vh_non-dissipated}\\
&v_c \approx \sqrt{\frac{4}{3}} (4 \pi)^{-1/5} (1-f_V)^{-1/5} \lambda^{1/10}  \left( \frac{\theta_0 L_j}{\rho_H t^2} \right)^{1/5} \label{eq:vc_non-dissipated}\\
&P_c \approx (4\pi)^{-2/5} (1-f_V)^{-2/5} \lambda^{1/5} \left( \frac{\theta_0^2 L_j^2 \rho_H^3}{t^4} \right)^{1/5} \label{eq:P_non-dissipated}\\
&R_j \approx \frac{(4 \pi)^{1/5}}{\sqrt{2 \pi}} (1-f_V)^{1/5} \lambda^{-1/10} {v_j}^{-1/2} \left( \frac{\theta_0^8 L_j^3 t^4}{{\rho_H}^3}\right)^{1/10} \label{eq:Rj_non-dissipated}
\end{align}
\end{subequations}
Notice that for $f_V \ll 1$, the cocoon velocity is very weakly dependent on $\lambda = \rho_{\rm eff}/\rho_H$ but the head velocity decreases with $\lambda$ as $\lambda^{-2/5}$. Therefore, we expect the head to become progressively slower in the presence of denser clouds.

It is straightforward to generalize the above scalings for the case when the jet is dissipated by clouds well within the cocoon (see section \ref{sec:pattern_const}).
In this case, the cocoon can be assumed to be a hemisphere with radius $r_c$. It is easy to check that in both the non-dissipated and dissipated 
regimes, the cocoon aspect ratio (approximated by $v_c/v_h$) is the same
\begin{equation}
\label{eq:vc_by_vh}
\frac{v_c}{v_h} \approx  \theta_0 \sqrt{\frac{\lambda}{3}} \,\approx \frac{\theta_{0,10}}{10}\: \sqrt{\lambda},
\end{equation}
so that even jets with a small opening angle can become dissipated in a medium with a sufficiently large density contrast, $\lambda$ (Eq. \ref{eq:lambda_def}), between the clouds and the diffuse medium. For a half-opening angle of $10^{\circ}$, this model predicts that the jet becomes dissipated for $\lambda \equiv \rho_{\rm eff}/\rho_H \gtrsim  100$. Using numerical simulations, in section \ref{sec:uniform_test} we show that a jet in a uniform medium indeed drives a self-similar cocoon+shock, but the ratio of cocoon to head speed is {$2\times$} larger than the value given by Eq. \ref{eq:vc_by_vh} (see Eq. \ref{eq:diss_ratio_app}), implying a {$4\times$} smaller density threshold $\lambda$ for dissipation. Basically, the cocoon properties scale as given by the analytic scalings, but the prefactors must be determined from numerical simulations (see section \ref{sec:results}).

\subsubsection{Uniformly distributed clouds}
\label{sec:uniform_clouds_analytic}
The simplest realization of a clumpy medium is a uniform medium with randomly distributed clouds of a given radius, $R_{\rm cl}$, {and a volume filling} fraction, $f_V$. Computing the effective density  
{that the jet-beam interacts with ($\rho_{\rm eff}$),} 
for a general distribution of these clouds (having density $\rho_{\rm cl}$) is challenging because of complex jet-cloud interaction. However, a physically motivated ansatz based on important dimensionless parameters of the problem for $\lambda \equiv \rho_{\rm eff}/\rho_H$ (assuming cold jet and medium) is
\begin{equation}
\label{eq:lambda_ansatz}
 \frac{\rho_{\rm eff}}{\rho_H} = \lambda \left( \frac{\rho_{\rm cl}}{\rho_H}, f_V, \frac{R_{\rm cl}}{R_j} \right),
\end{equation}
where the precise form of this function is unknown but will be calibrated with simulations in section \ref{sec:jets_clumpy}. The dependence on jet properties in Eq. \ref{eq:lambda_ansatz} goes into the jet radius $R_j$.
Note that we can construct more dimensionless parameters, but $\rho_{\rm cl}/\rho_H$, $f_V$ and $R_{\rm cl}/R_j$\footnote{Cloud radius $R_{\rm cl}$ can be normalized with other length-scales such as $r_c$, $z_h$, but the jet velocity (see Eq. \ref{eq:Tint}) and its radius should affect jet-cloud interaction more directly.} appear to be most relevant. In our simulations with variation of $R_{\rm cl}$, we find that $\lambda$ is rather insensitive to $R_{\rm cl}$. However, this ansatz needs to be tested more closely with further numerical simulations.

We can further simplify things by separating the dependence on cloud density ($\rho_{\rm cl}$) and the cloud pattern ($f_V$, $R_{\rm cl}/R_j$). For this, we model the head to be obstructed by a medium of density $\rho_{\rm cl}$ (clouds) for a fraction $f_L$ of the path length, and $\rho_H$ for a fraction $(1-f_L)$ of the path length. Consider a column of length $L$ which is large compared to the separation between clouds but small compared to the total length of the jet, $z_h$. The time taken to traverse this distance (using Eq.  \ref{eq:v_h}) is given by
\begin{equation}
\begin{split}
\label{eq:Tint}
&T = \left( \frac{L}{v_j \sqrt{\rho_j}} \right) \left( f_L \sqrt{\rho_{\rm cl}} + (1 - f_L) \sqrt{\rho_H} \right) \\
&\ \ \equiv \left( \frac{L}{v_j \sqrt{\rho_j}} \right) \sqrt{\rho_{\rm eff}}.
\end{split}
\end{equation}
This expression gives the required expression for $\lambda = \rho_{\rm eff}/\rho_H$ as
\begin{equation} 
\label{eq:lambda_fL_rho}
\lambda = \left( f_L \sqrt{\frac{\rho_{\rm cl}}{\rho_H}} + (1-f_L)  \right)^2,
\end{equation}
so that the unknown cloud length fraction, $f_L$, is only a function of the spatial distribution and size of the clouds. That is, for our setup
\begin{equation} \label{eq:f_L_def}
f_L \equiv f_L \left(f_V, \frac{R_{\rm cl}}{R_j}\right).
\end{equation}
Here, we assume that the only relevant dimensionless parameters determining $f_L$ are cloud volume fraction ($f_V$) and the ratio of the cloud and jet radii ($R_{\rm cl}/R_j$). Thus, we have separated the dependence of $\lambda$ on density contrast ($\rho_{\rm cl}/\rho_H$), and cloud-size ($R_{\rm cl}$) and distribution ($f_V$).

Although a simple length to volume relation should be $f_L \propto f_V^{1/3}$ 
(independent of $R_{\rm cl}$), 
the calculation of $f_L$ for any given cloud pattern from first principles is still difficult due to several reasons. For example, there are large uncertainties associated with modeling the interaction between a jet and a single cloud such as, the distance over which a single cloud gets dragged before being destroyed and the contribution due to the overlap of wakes of different clouds are not known accurately in general. In some cases, the expanding {forward shock} could pre-shock and destroy a cloud partially before the jet head interacts with it. Such processes depend on the cloud radius, $R_{\rm cl}$ which could be a parameter deciding $f_L$ (and hence the assumed $R_{\rm cl}$ dependence). To make further progress, we use numerical simulations with varying cloud patterns (holding the cloud density fixed) to calibrate the functional dependence of $f_L$ on $f_V$ and $R_{\rm cl}/R_j$  in section \ref{sec:rho_const}.

\begin{figure*}
\begin{centering}
\includegraphics[width=0.5\textwidth, clip=true, trim={0 0 0 0.8cm}]{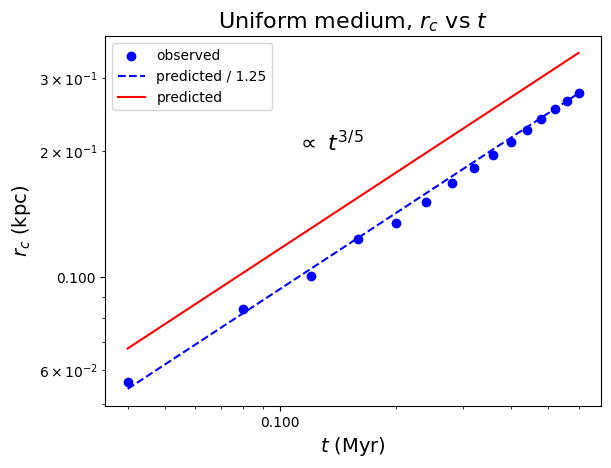}
\includegraphics[width=0.483\textwidth, clip=true, trim={0 0 0 0.8cm}]{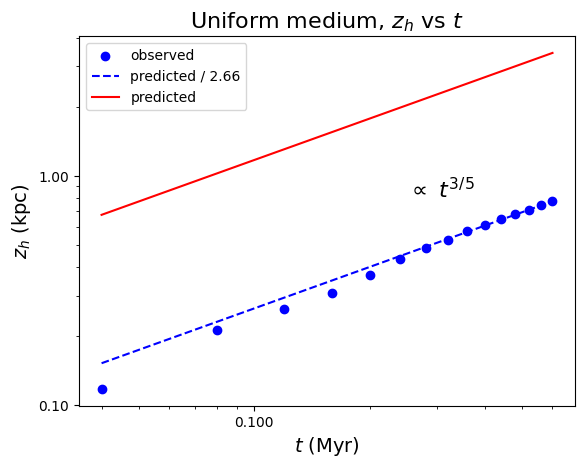}
\end{centering}
\caption{Blue dots show the measured values of cocoon radius $r_c$ ({left} panel) and $z_h$ ({right} panel) at different times for our uniform run (see Table \ref{table:rho_const}). Red lines show the predicted values from the analytical model (integrating Eqs. \ref{eq:vc_uniform_pred} and  \ref{eq:vh_uniform_pred}). Blue dashed lines show the predicted values scaled to match the observed values at $t \approx 0.6$ Myr.} 
\label{fig:uniform_vs_t}
\end{figure*}

\section{Results}
\label{sec:results}

In this section, we present the results from our numerical simulations of jets in a uniform and clumpy medium and compare them with analytic estimates obtained in section \ref{sec:analytic_estimates}. Firstly, we compare the analytical predictions using a simulation of a jet propagating in a uniform medium (section \ref{sec:uniform_test}). 
We find that while the scaling with time of cocoon radius and jet-head location follow the analytic estimates, the exact prefactors are different.
Following this, we verify our model using analytic scalings with jet simulations in a clumpy medium. However, due to the uncertainties in estimating the geometrical length factor $f_L$ from first principles, in section \ref{sec:rho_const} we first perform simulations keeping cloud density constant and varying the pattern of clouds (i.e., their $f_V$ and $R_{\rm cl}$) to study the dependence of $f_L$ on $f_V, R_{\rm cl}$. Using the $f_L$ value inferred from these simulations, we finally test the predictions of our calibrated model for the scaling of the head velocity $v_h$ with the effective density encountered by the jet ($\rho_{\rm eff}$) in section \ref{sec:pattern_const}  by choosing a fixed cloud pattern (fixed $f_L$) and varying the cloud density ($\rho_{\rm cl}$).

\subsection{Analytical model versus simulations for a uniform medium}
\label{sec:uniform_test}
Here we compare the analytic model of propagation of a non-relativistic jet in a uniform medium (Eqs. \ref{eq:vh_uniform_pred} and \ref{eq:vc_uniform_pred}) with results from numerical simulation, using the jet/ISM parameters listed in Table \ref{table:jet_sim_parameters}
and a cubic box of side $L_{\rm box} = 0.8$ kpc. We are able to follow the evolution of the jet upto a time of 0.6 Myr, after which the jet-head exits the domain (see Fig. \ref{fig:uniform_end}). 

The time evolution of the cocoon lateral size ($r_c$) and the jet-head position ($z_h$) are shown in Fig. \ref{fig:uniform_vs_t}. The predicted values for these are obtained by integrating Eqs. \ref{eq:vc_uniform_pred} and \ref{eq:vh_uniform_pred} from 0 to $t$. For both the head and the cocoon, the predicted scaling with time of $r_c$, $z_h$ $\propto t^{3/5}$ is in good agreement with the simulation results. However, the predicted numerical pre-factors are larger by a factor of 1.25 for $r_c$, and 2.66 for $z_h$. There are several possible reasons for this discrepancy. The jets are not purely kinetic energy driven in the simulations. Figure \ref{fig:uniform_end} shows that after the collimation shock, the jet is no longer a laminar type flow and rather contains turbulent structures. Such structures reduce the directional ram pressure of the jet which results in a smaller height for the jet head. Moreover, the jet contains significant thermal energy after collimation (see middle panel of fig \ref{fig:uniform_end}), in contrast to the assumption of a cold jet in the analytical model. The expressions (Eqs. \ref{eq:vh_uniform_pred}-\ref{eq:Rj_uniform_pred}) should have prefactors that depend on the gas adiabatic index ($\gamma$). Additionally, the geometries of the jet and the cocoon outer shock, which we have assumed to be cylindrical in the analytical model, are not so in simulations. Also, the model assumes a self-similar solution at all times, but this is likely to be broken at early times, e.g., when the jet collimation has not yet been completed. The model does not account for this early phase of jet launching. Previous studies of relativistic jets (\cite{harrison2018}) have highlighted the need for calibrating the analytical models of jet propagation derived from \cite{bromberg11} via numerical simulations.

Using the calibration obtained from our jet simulation in a uniform medium, the ratio of cocoon size to the jet vertical extent is found to be
\begin{equation} 
\label{eq:diss_ratio_app}
\frac{r_c}{z_h} = \frac{v_c}{v_h} \approx 1.23\ \theta_0 \approx 2.15 \times \frac{\theta_{0, 10}}{10}, 
\end{equation}
assuming that the scaling of this ratio with $\theta_0$ remains unchanged (see Eq. \ref{eq:vc_by_vh} with $\lambda=1$). Further assuming that the scaling of this ratio with the density contrast, $\lambda$ (Eq. \ref{eq:lambda_def}), is  unchanged, this ratio in a clumpy medium should be 
\begin{equation}
    \frac{r_c}{z_h} = 2.15 \times \frac{\theta_{0, 10}}{10} \: \sqrt{\lambda}\,.
\end{equation}
Therefore, we obtain {a} prediction for the critical value of $\lambda$ required for jet dissipation 
(i.e. $r_c = z_h$), 
\begin{equation} 
\label{eq:calibrated_lambda_crit}
\lambda_{\rm crit} = \frac{21.7}{\theta_{0, 10}^2},
\end{equation}
{above which the jet will be dissipated.}

Fig. \ref{fig:uniform_end} shows the snapshots of the density (left panel), pressure (middle panel) and vertical velocity (right panel) for our uniform medium run. Clearly, the cocoon geometry is far from cylindrical (cocoon shape depends on background density profile; e.g., see Fig. 3 in \citealt{harrison2018}) and $z_h/r_c$ is smaller than $\sqrt{3}/\theta_0 \approx 10$, the value expected from self-similar theory (see Eq. \ref{eq:vc_by_vh} with $\lambda =1$). This small 
{discrepancy} is also observed in earlier relativistic simulations (\citealt{harrison2018}), but 
Fig. \ref{fig:uniform_vs_t} shows that the evolution 
follows the same time-dependence as estimated in the analytical theory (section\ref{sec:uniform_analytic}).

In Fig. \ref{fig:uniform_end}, one can clearly see the jet collimation shock, an almost cylindrical jet-beam with a turbulent surface, a high velocity backflow in the inner cocoon, and a dense outer cocoon with Kelvin-Helmholtz vortices at the inner edge.
The pressure is roughly uniform in the cocoon because of the high temperature and the associated large sound speed compared to the flow speeds. Most galaxy formation simulations, with resolutions insufficient to resolve the jet radius (see Eq. \ref{eq:Rj_uniform_pred}) cannot capture, even qualitatively, the evolution of the cocoon and energy dissipation by AGN jet feedback (see section \ref{sec:jets_cosmo} for further discussion).

\subsection{Jets in a clumpy medium}
\label{sec:jets_clumpy}

\begin{figure*}
\includegraphics[width=0.33\textwidth]{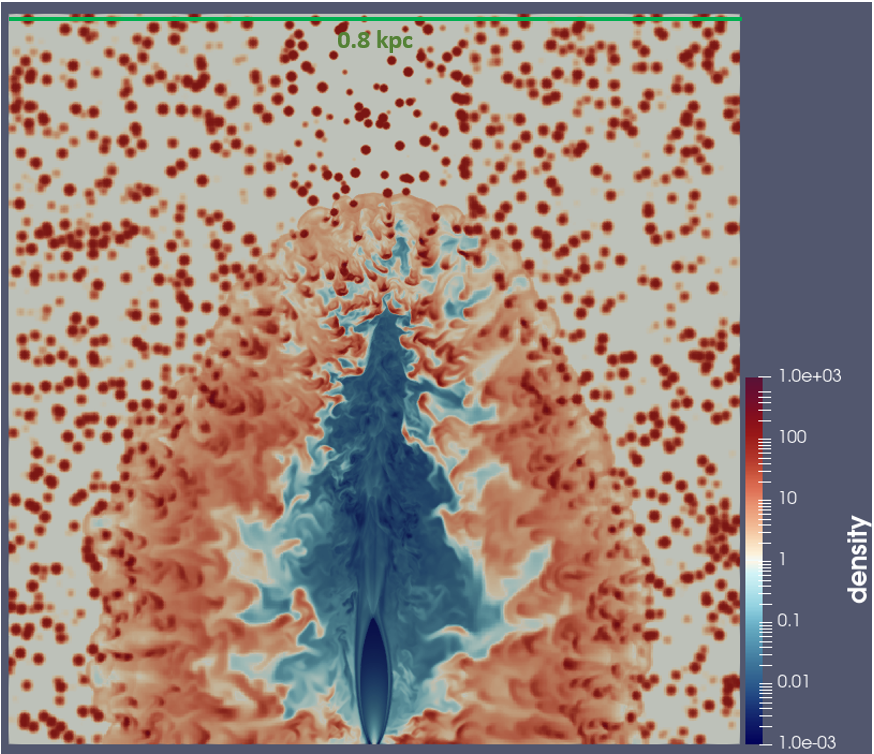}
\includegraphics[width=0.33\textwidth]{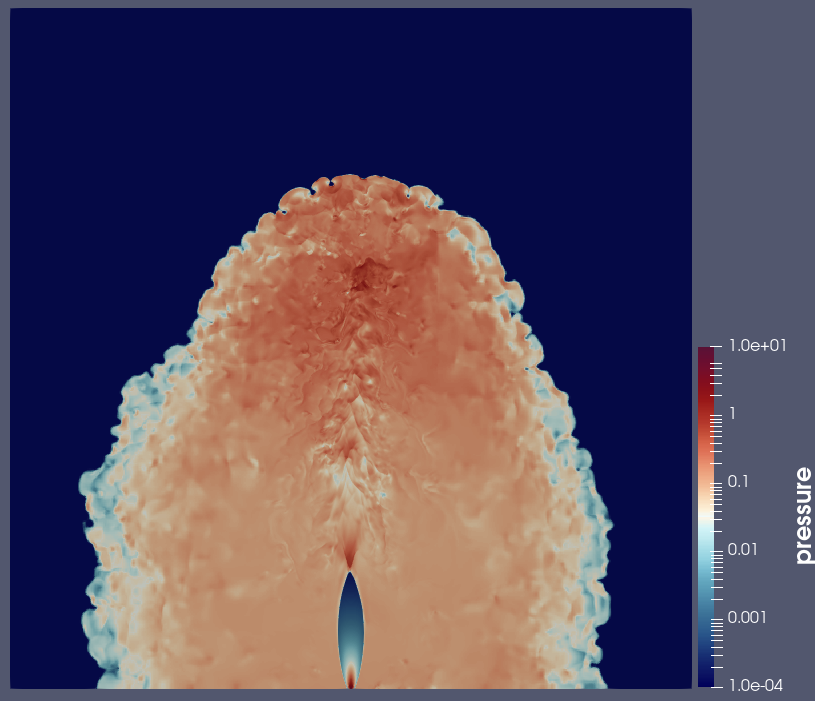}
\includegraphics[width=0.33\textwidth]{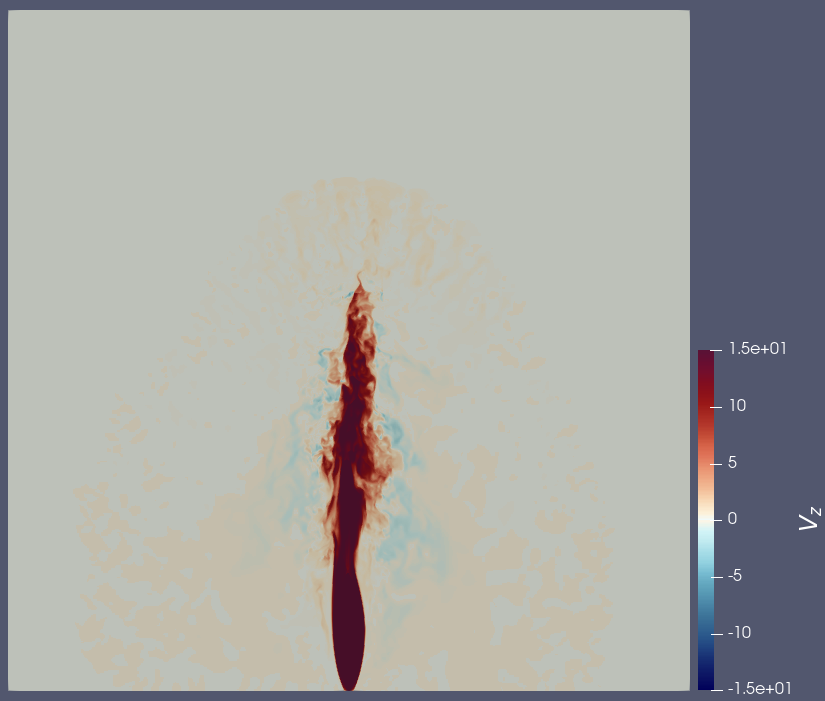}
\caption{
{2D slices (y-z plane)} of density ($m_p$ cm$^{-3}$, left), pressure ($1.67 \times 10^{-8}\ {\rm dyn~cm}^{-2}$, center) and vertical velocity (1000 km s$^{-1}$, right) for jet in a medium with $\rho_H = 1\ m_p$ cm$^{-3}$, $\rho_{\rm cl} = 100\ m_p$ cm$^{-3}$ and $f_V = 0.1$, at time $t = 0.96$ Myr. Note that even though many of the clouds inside the cocoon are largely intact 
({since} destruction time inside the cocoon is long),  they have already come in pressure equilibrium with the cocoon as the shock driven by the high-pressure cocoon can compress these clouds on timescales $\sim 0.1\ {\rm Myr}$ (see text around Eq. \ref{eq:t_s_cl}). The backflows with clouds are weaker than in the uniform case shown in Fig. \ref{fig:uniform_end}.
}  
\label{fig:fiducial}
\end{figure*}

Clumps affect the cocoon scalings because of the different response of the jet and cocoon to clouds, as discussed in section \ref{sec:jet_clumpy_analytic}. The expressions for the head and cocoon velocities in a clumpy medium (Eqs. \ref{eq:vh_non-dissipated}, \ref{eq:vc_non-dissipated}) and in a uniform medium (Eqs. \ref{eq:vh_uniform_pred}, \ref{eq:vc_uniform_pred}) have a very similar form, but have additional dependence on $\lambda = \rho_{\rm eff}/\rho_H$ and a weak dependence on $f_V$. Here, $\lambda$ itself depends on $\rho_{\rm cl}/\rho_H$, $f_V$ and $R_{\rm cl}/R_j$ (Eq. \ref{eq:lambda_ansatz}), which can be further decomposed into the dependence on the cloud density contrast ($\rho_{\rm cl}/\rho_H$) and the cloud pattern ($f_V$, $R_{\rm cl}/R_j$; see Eqs. \ref{eq:lambda_fL_rho}, \ref{eq:f_L_def}). Now, through numerical simulations, we decouple the dependence on the cloud pattern (section \ref{sec:rho_const}) and on the cloud density (section \ref{sec:pattern_const}).

{\em Important timescales for clouds within the cocoon:} Before describing results from simulations of jets interacting with a clumpy ISM, we estimate some timescales that help us understand the dynamics of clouds within the cocoon not directly interacting with the jet-beam. Note that the sound crossing time across the clouds  
\begin{equation}
\label{eq:t_s_cl}
t_{s,{\rm cl}} \equiv \frac{2 R_{\rm cl}}{c_{s,{\rm cl}}} = 6.7~{\rm~Myr}~R_{\rm cl, 5} T_{H,4}^{-1/2} \sqrt{\frac{\rho_{\rm cl}/\rho_H}{100}}, 
\end{equation}
where $R_{\rm cl,5}$ is the cloud radius normalized to 5 pc and $T_{H,4}$ is the temperature of the homogeneous ISM in units of $10^4$ K. Since this timescale is much longer than the dynamical time, the clouds are expected to be isochoric and under-pressured just after the passage of the forward shock. Being under-pressured, clouds are expected to implode on a timescale $t_{\rm imp} \sim R_{\rm cl}/\sqrt{P_c/\rho_{\rm cl}} \sim t_{s,{\rm cl}} (P_H/P_c)^{1/2} \sim 0.02$ Myr, much shorter than the sound crossing time ($P_H$ is the ISM pressure). Simultaneously, clouds are being mixed by shear instabilities on $t_{\rm cc} \sim R_{\rm cl}(\rho_{\rm cl}/\rho_c)^{1/2}/v_c$ timescale, which can be longer than the implosion time by $\sim \sqrt{\rho_H/\rho_c}$ within the inner cocoon (assuming the relative velocity everywhere within the cocoon $\sim v_c$). Thus, the clouds embedded in the hot cocoon are compressed, mixed and pushed (due to the motion of cocoon material relative to the clouds). Also note that the cocoon velocity and its pressure are changing on a dynamical time (Eqs. \ref{eq:P_uniform_pred}, \ref{eq:vc_uniform_pred}). Because of these various complexities, it is difficult to predict the cloud evolution within the cocoon quantitatively.

Fig. \ref{fig:fiducial} shows the snapshots of density, pressure and vertical velocity for our fiducial clumpy run f01R5. The dense shell is much thicker because of mixing of clouds into the shell. The clouds near the outer periphery of the dense shell are under-pressured but become isobaric inside the cocoon, consistent with the estimate in the previous paragraph. The backflow in the inner cocoon is irregular and less prominent. Most importantly, the cocoon is  more spherical compared to the one in a uniform medium shown in Fig. \ref{fig:uniform_end}.

\begin{figure*}
\begin{centering}
%\begin{tabular}{c}
{\mbox{\includegraphics[width=0.5\textwidth]{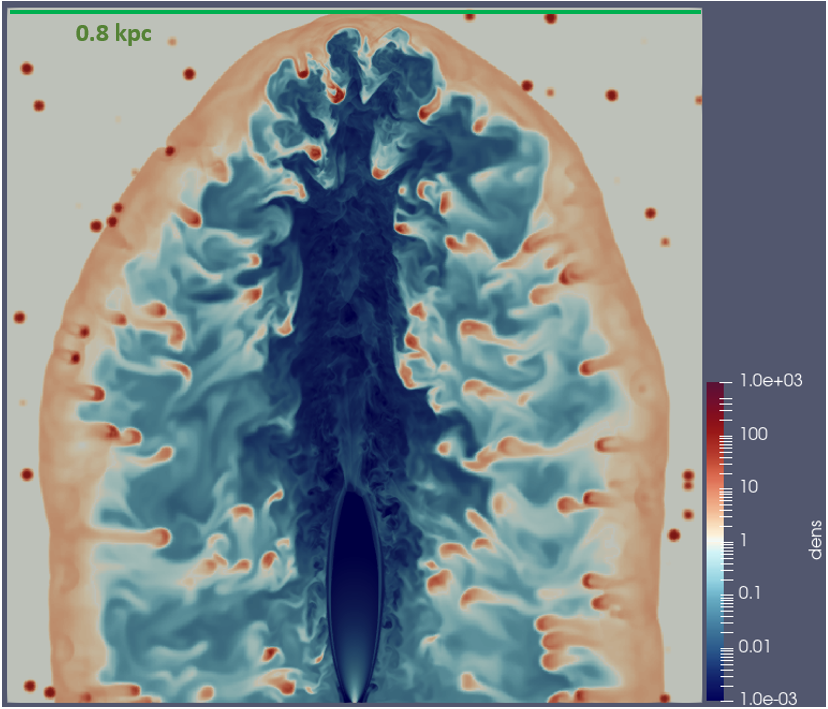}}}%\\
{\mbox{\includegraphics[width=0.495\textwidth]{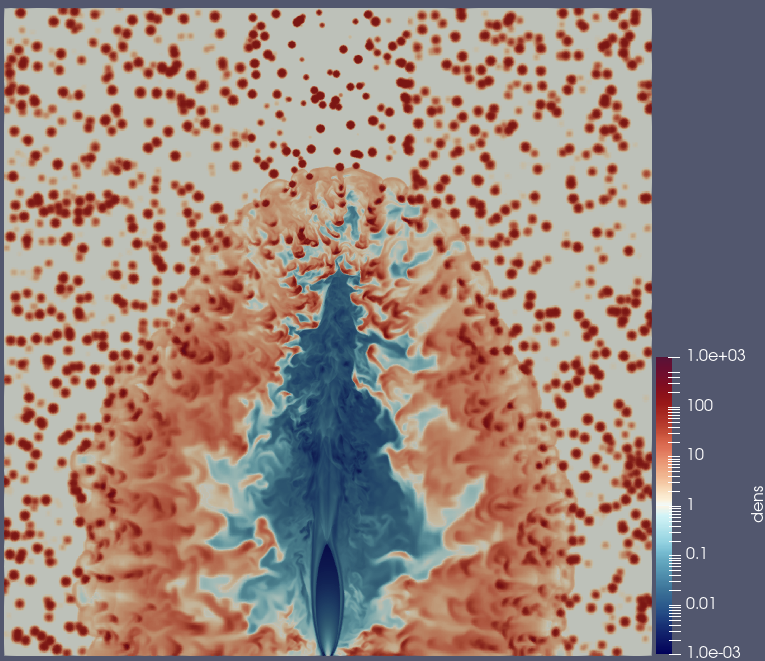}}}%\\
%\end{tabular}
\end{centering}
\caption{Snapshots in y-z plane of density for cloud density of 100 $m_p\ {\rm cm}^{-3}$, $R_{\rm cl}=5$ pc and $f_V=0.00625$ (left panel) and $f_V=0.1$ (right panel). As expected, the cocoon is more spherical and dissipated for a higher volume fraction of clouds.} 
\label{fig:vary_Ncl}
\end{figure*}

\subsubsection{Effect of changing cloud pattern ($f_V$, $R_{\rm cl}$)} 
\label{sec:rho_const}

\begin{figure*}
\begin{centering}
%\begin{tabular}{c}
{\mbox{\includegraphics[width=0.5\textwidth]{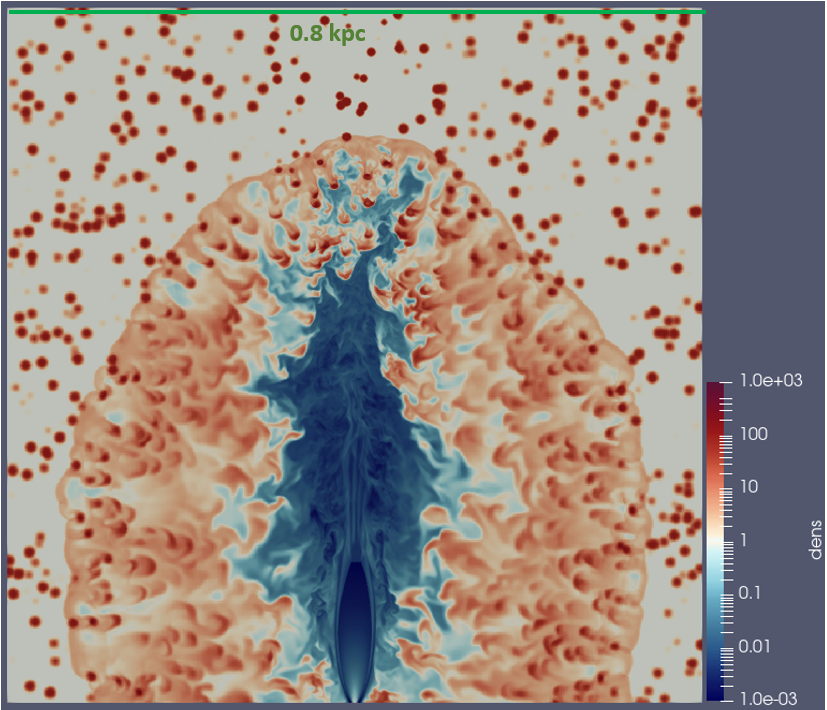}}}%\\
{\mbox{\includegraphics[width=0.5\textwidth]{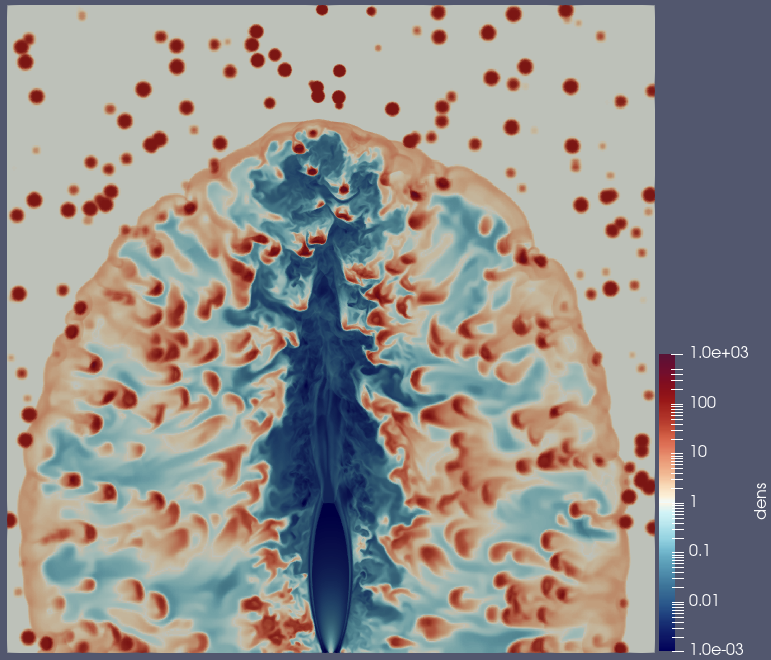}}}%\\
%\end{tabular}
\end{centering}
\caption{Snapshots of density in y-z plane for cloud density of 100 $m_p\ {\rm cm}^{-3}$, $f_V=0.05$ and $R_{\rm cl}=5$ pc (left panel) and $R_{\rm cl}=7.9$ pc (right panel). The two cocoons have a similar aspect ratio (see Fig. \ref{fig:fV_vh_vc}) and are not dissipated. Smaller clouds (left panel) are disrupted more compared to larger ones (right panel) and mass load the shocked ISM.} 
\label{fig:vary_Rcl}
\end{figure*}

In this section, we present runs from which we can deduce the dependence of the effective length fraction of clouds encountered by the jet ($f_L$) on the cloud volume fraction ($f_V$) and the cloud-size ($R_{\rm cl}$; see Table \ref{table:rho_const} for the list of these runs), before exploring the $\rho_{\rm cl}/\rho_H$
dependence in section \ref{sec:pattern_const}. 
Therefore, here we study the effect of changing the number and size of clouds (keeping $\rho_{\rm cl}$ constant at 100 $m_p$ ${\rm cm}^{-3}$) on the head and cocoon dynamics. For this, we perform simulations with a wide range of volume filling fractions ranging from 0.00625 to 0.1, and cloud radii between 5 and 10 pc. In order to avoid statistical fluctuations because of the random positioning of clouds across these runs, we fix individual cloud positions. Some clouds are then removed to reduce $f_V$ and/or to increase $R_{\rm cl}$ (see Fig. \ref{fig:initial_hollowed}). Snapshots of density for simulations with constant $R_{\rm cl}$ but varying $f_V$ are shown in Fig. \ref{fig:vary_Ncl}, while those with constant $f_V$ but varying $R_{\rm cl}$ are shown in Fig. \ref{fig:vary_Rcl}. The figures indicate that the volume filling fraction significantly affects the cocoon size and shape, whereas the cloud radius only has a mild effect. Therefore, one can qualitatively expect that $f_L$ (and therefore $\lambda$) does not have a significant dependence on the cloud size (as long as $f_V$ remains the same).
In all cases, the cocoon engulfs the clouds, which are partially disrupted by the dense shell but are not much affected once inside the lower-density inner cocoon. The smaller clouds (due to a shorter cloud-crushing time) and clouds with larger volume filling fraction (since material from many clouds is shredded) mass load the shocked ISM shell.

\begin{table} 
\begin{center}
\scalebox{0.87}{
\hspace*{-1.7cm}\begin{tabular}{c c c c c c}
\hline
Label & $R_{\rm cl}$  & $f_V$ & $v_h$ & $v_c$ & $f_L$\\
 & (pc) &  & ($10^3$ km s$^{-1}$) & ($10^3$ km s$^{-1}$) & \\
\hline \hline
uniform & 0 & 0.0 & 0.99 & 0.33 & 0.0\\
\hline
f0006R5 & 5 & 0.00625 & 0.64 & 0.33 & 0.08\\
\hline
f0012R5 & 5 & 0.0125 & 0.57 & 0.33 & 0.11\\
\hline
f0025R5 & 5 & 0.025 & 0.45 & 0.32 & 0.19\\
\hline
f005R5 & 5 & 0.05 & 0.41 & 0.30 & 0.22\\
\hline
f0075R5 & 5 & 0.075 & 0.39 & 0.29 & 0.24\\
\hline
f01R5 & 5 & 0.1 & 0.36 & 0.28 & 0.28\\
\hline
f0012R6.3 & 6.3 & 0.0125 & 0.50 & 0.33 & 0.15\\
\hline
f0025R6.3 & 6.3 & 0.025 & 0.44 & 0.33 & 0.20\\
\hline
f0025R7.9 & 7.9 & 0.025 & 0.43 & 0.33 & 0.21\\
\hline
f005R7.9 & 7.9 & 0.05 & 0.43 & 0.34 & 0.21\\
\hline
f01R10 & 10 & 0.1 & 0.29 & 0.34 & 0.41\\
\hline
\end{tabular}
}
\end{center}
\caption{\label{table:rho_const}List of simulations where the cloud density is held constant at 100 $m_p$ ${\rm cm}^{-3}$ and $f_V$, $R_{\rm cl}$ are varied. The simulation name fxxxRy indicates a volume filling fraction, $f_V = x.xx$ and a cloud radius, $R_{\rm cl} = y$ pc. For these runs, the simulation box is a cube of side 0.8 kpc, with the highest resolution on the jet axis being 0.4 pc. Velocities are averaged over the time when the jet material is still in the box. 
}
\end{table}

\begin{figure}
\centering
% \begin{centering}
% \begin{tabular}{c}
% {\mbox{\includegraphics[width=8cm]{results_plots/fV_vh_vc.png}}}
\includegraphics[width=0.45\textwidth]{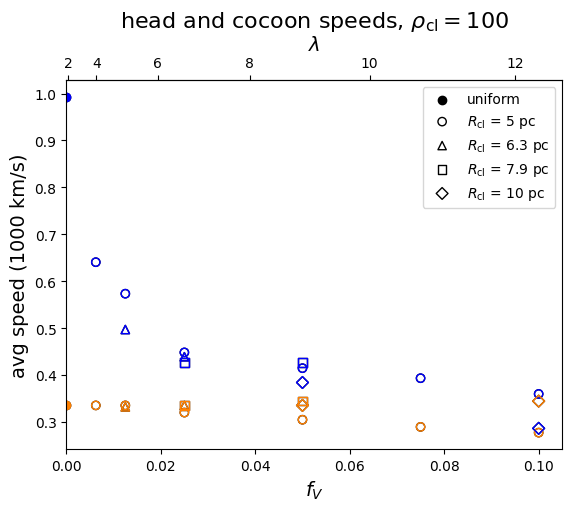}
% \end{tabular}
% \end{centering}
\caption{Variation of the average head velocity ($v_h$, blue) and average cocoon velocity ($v_c$, orange) as a function of $f_V$ for $\rho_{\rm cl}$ = 100 $m_p$ ${\rm cm}^{-3}$. Different symbols correspond to different cloud sizes $R_{\rm cl}$. For $f_V \gtrsim 0.02$, 
the average head speed decreases very slowly with $f_V$. The $\lambda$ values shown on top are derived from Eq. \ref{eq:lambda_fV_relation}.} 
\label{fig:fV_vh_vc}
\end{figure}

\begin{figure}
\centering
\includegraphics[width=0.45\textwidth]{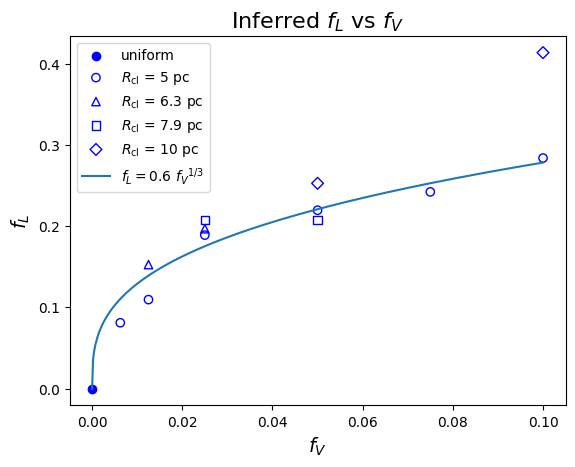}
\caption{Variation of the effective cloud path length $f_L$ (as inferred from our numerically calibrated analytical model, Eq. \ref{eq:fL-fV-analytical}) 
as a function of $f_V$, for the same set of simulations shown in Fig. \ref{fig:fV_vh_vc}.  The solid line shows a fit $f_L = 0.6 f_V^{1/3}$ that matches the simulation results; $R_{\rm cl}$ = 10 pc and $f_V=0.1$ run may be an outlier due to very few clouds along the jet-beam.}
\label{fig:fL_vs_fV}
\end{figure}

This set of simulations is designed to capture the transition from the highly non-dissipated regime (very few or no clouds) to the near-dissipated regime (large $f_V$). The cocoon expansion velocity in Fig. \ref{fig:fV_vh_vc} shows very little variation across all runs as expected (from Eq. \ref{eq:vc_non-dissipated}
), and stays at around 300 ${\rm km} {\rm s}^{-1}$. Fig. \ref{fig:fV_vh_vc} shows that the head velocity decreases rapidly with increasing $f_V$ for small $f_V$, then saturates (or decreases very slowly) at around 400 ${\rm km} {\rm s}^{-1}$ for $f_V \gtrsim 0.02$. Note that none of the runs in Table \ref{table:rho_const}, except for $R_{\rm cl} = 10$ pc and $f_V=0.1$ for which $v_h < v_c$ (see Fig. \ref{fig:fV_vh_vc}), are fully dissipated since $v_h > v_c$. However, for $f_V \gtrsim 0.02$ there is no prominent jet head and there are at least a few clouds obstructing the jet-beam within the vertical extent of the cocoon (right panel of Fig. \ref{fig:vary_Ncl} and Fig. \ref{fig:vary_Rcl}). There is a continuous decrease in the average head velocity with increasing $f_V$, before it decreases below the cocoon's lateral expansion velocity and the jet dissipates. The figure also shows the simulations with the same $f_V$ but different cloud radii. We find that the cloud radius does not have much effect on the cocoon dynamics, and $f_V$ is the important parameter governing $f_L$ and $\rho_{\rm eff}$.

Now, based on the head velocities measured here and the equations of the analytical model (Eqs. \ref{eq:vh_non-dissipated}, \ref{eq:lambda_fL_rho}), we can compute a value of $f_L$ for each of these simulations. For this, we note that in the case of uniform medium, $\lambda=1$. Then using Eq.  \ref{eq:vh_non-dissipated} (neglecting the $(1-f_V)$ dependence as $f_V \ll 1$), we obtain
\begin{equation}
\lambda = \frac{\rho_{\rm eff}}{\rho_H} = \left(\frac{v_h}{v_{h, {\rm uniform}}}\right)^{-5/2},
\end{equation}
which gives us the required $\rho_{\rm eff}$ for several combinations of $f_V$ and $R_{\rm cl}$. Substituting this $\rho_{\rm eff}$ in Eq.  \ref{eq:lambda_fL_rho}, we can solve for $f_L$ as
\begin{equation}
f_L = \frac{\sqrt{\rho_{\rm eff}} - \sqrt{\rho_H}}{\sqrt{\rho_{\rm cl}} - \sqrt{\rho_H}}
\label{eq:fL-fV-analytical}
\end{equation}
A plot of the $f_L$ values thus obtained as a function of $f_V$  is shown in Fig. \ref{fig:fL_vs_fV}. We observe that $f_V$ is the major factor in controlling $f_L$, changing $R_{\rm cl}$ at the same $f_V$ does not change $f_L$ significantly.\footnote{While we have varied $f_V$ by more than two orders of magnitude, we could vary the cloud size only within a factor of 2. Smaller clouds become numerically unresolved and with larger clouds the number of clouds along the jet-beam is small and there is a large statistical variation in simulation outcomes. The length fraction $f_L$ may depend on $R_{\rm cl}$ for a larger range of variation, but we do not explore this here.} Except for ($f_V,R_{\rm cl}$) = (0.1, 10 pc) run, which appears to be an outlier, 
there is no systematic variation of the head and cocoon speeds with the cloud size.
This outlier value may be due to the small number of clouds along the path of the jet.
In Fig. \ref{fig:fL_vs_fV}, $f_L$ increases sharply from 0 with increasing $f_V$, then grows slowly to values of $\approx 0.2 - 0.3$ for $f_V \gtrsim 0.02$. For example, for $\left(f_V, R_{\rm cl}\right) = \left(0.1, 5\ {\rm pc}\right)$, $f_L$ is found to be 0.28.
{W}e find that a scaling
\begin{equation} 
\label{eq:f_Lf_V_relation}
f_L = 0.6\ {f_V}^{1/3}
\end{equation}
fits the data quite well (solid line in Fig. \ref{fig:fL_vs_fV}), 
as expected. The normalization factor is, however, non-trivial to obtain from first principles since it involves complex jet-cloud interactions as noted in section \ref{sec:analytic_estimates}. 
Of course, this scaling (and other assumptions of our model) will break down as $f_V$ approaches unity (and $f_L$ approaches 0.6). 
Now, 
for $\rho_{\rm cl}/\rho_H = 100$, 
and $f_V = 0.1$, 
$f_L = 0.6 f_V^{1/3} = 0.28$, 
{and } Eq. \ref{eq:lambda_fL_rho} gives $\lambda = 12.4$ which is
$< \lambda_{\rm crit}$.
Thus, none of the simulations in Table \ref{table:rho_const}  
(except f01R10 which has a large outlier $f_L=0.41$)  
is expected to be fully dissipated, consistent with Fig. \ref{fig:fV_vh_vc}.

\begin{figure*}
\begin{centering}
{\mbox{\includegraphics[width=0.5\textwidth]{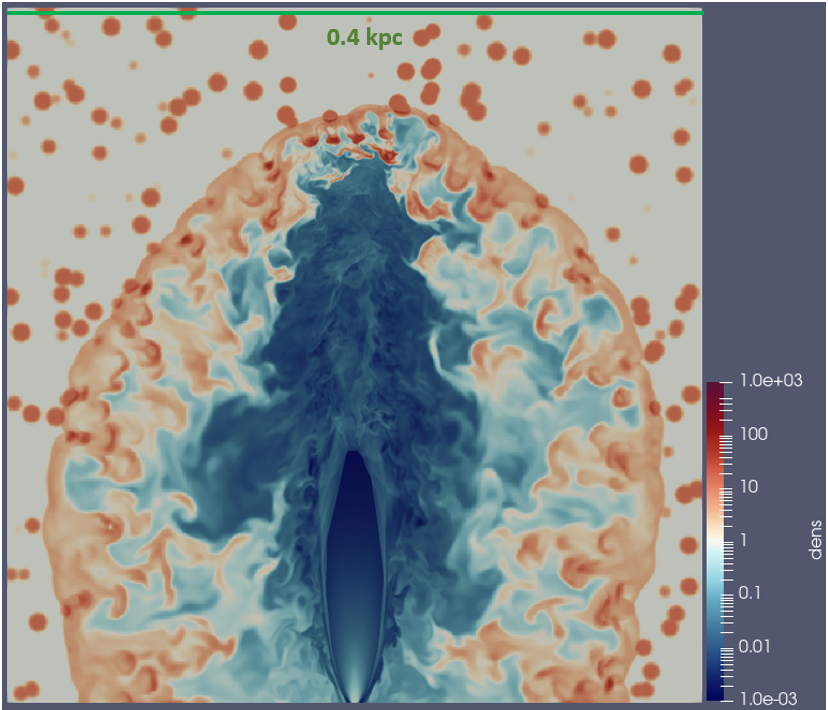}}} 
{\mbox{\includegraphics[width=0.5\textwidth]{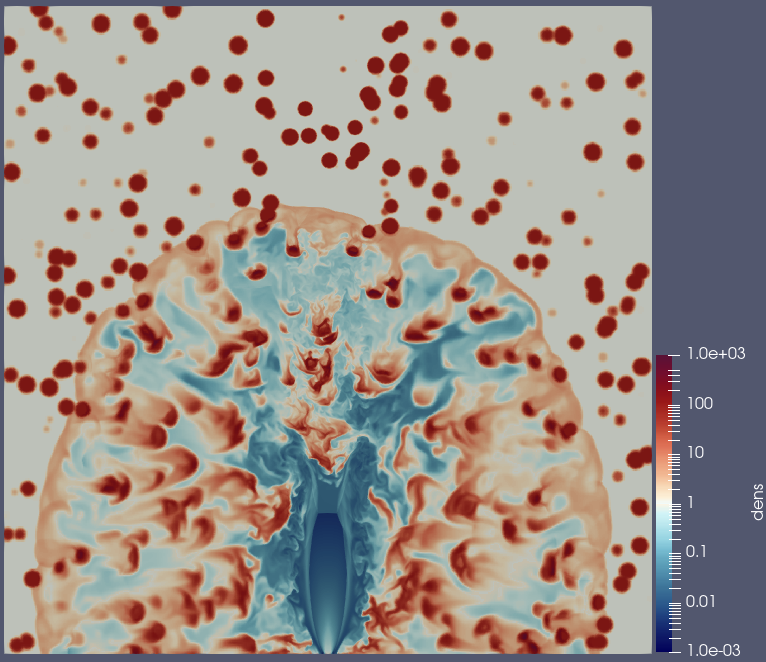}}}
\end{centering}
\caption{Snapshots of density for $f_V=0.1$, and cloud densities of 12 $m_p\ {\rm cm}^{-3}$ ($\lambda = 2.9$) (left panel) and 100 $m_p\ {\rm cm}^{-3}$ ($\lambda = 12.6$) (right panel), which represent the non-dissipated and fully-dissipated regimes respectively. For the former jet power dissipates at the clouds close to the vertical cocoon boundary, whereas for the latter the jet-beam is clearly obstructed by clouds well within the cocoon.} 
\label{fig:clumpy_fV_01}
\end{figure*}

\subsubsection{Effect of changing cloud density ($\rho_{\rm cl}$)}
\label{sec:pattern_const}
With the values of the effective length fraction covered in clouds ($f_L$) inferred from section \ref{sec:rho_const}, we are now in a position to determine $\rho_{\rm eff}$ for a given pattern ($f_V,~R_{\rm cl}$) and density of clouds ($\rho_{\rm cl}$). Thus, we can now test the predicted scaling of $v_h$ and the ratio $\lambda$ ($\rho_{\rm eff}/\rho_H$), by fixing a particular cloud pattern (with $f_L$ inferred from Eq. \ref{eq:f_Lf_V_relation}) and changing the density of the cloud material. In this section, we present the results for jet-head and cocoon dynamics from the simulations where the distribution of clouds is held fixed at $\left(f_V, R_{\rm cl}\right) = \left(0.1, 5\ {\rm pc}\right)$ and cloud density is varied from 12 $m_p\ {\rm cm}^{-3}$ to 100 $m_p\ {\rm cm}^{-3}$ (see Table \ref{table:pattern_const}). For this cloud distribution, we find $f_L$ to be 0.284 in section \ref{sec:rho_const} for $\rho_{\rm cl}=100 m_p$ cm$^{-3}$, and we assume the same value of $f_L$ for all cloud densities. We use this value of $f_L$ to compute $\rho_{\rm eff}$ corresponding to each value of $\rho_{\rm cl}$, using Eq.  \ref{eq:lambda_fL_rho}. 
Snapshots of density for the two extreme cases ($\rho_{\rm cl} = 12,~100 m_p$ cm$^{-3}$ 
or, $\lambda = 2.9, 12.65$) 
are shown in Fig. \ref{fig:clumpy_fV_01}.

\begin{table}
\begin{center}
\scalebox{0.87}{
\hspace*{-1.55cm}\begin{tabular}{c c c c c}
\hline
Label & $\rho_{\rm cl}$ & $v_h$ & $v_c$  & Dissipated \\
 & ($m_p\ {\rm cm}^{-3}$) & ($10^3$ km s$^{-1}$) &  ($10^3$ km s$^{-1}$) &  \\
\hline \hline
f01d12 & 12 & 1.00 & 0.42 & No\\
\hline
f01d25 & 25 & 0.74 & 0.42 & No\\
\hline
f01d50 & 50 & 0.65 & 0.41 & No\\
\hline
f01d75 & 75 & 0.42 & 0.34 & Marginally\\
\hline
f01d100 & 100 & 0.16 & 0.36 & Yes\\
\hline
\end{tabular}
}
\end{center}
\caption{\label{table:pattern_const}List of simulations where the pattern of clouds is held constant at $R_{\rm cl} = 5$ pc, $f_V = 0.1$ and cloud density ($\rho_{\rm cl}$) is varied. For these runs, the simulation box is a cube of side 0.4 kpc and the highest resolution is 0.2 pc. Velocities are averaged over the time when the jet has collimated and jet material is still in the box. Unlike for constant $\rho_{\rm cl}$ runs in Table \ref{table:rho_const}, we do not hollow out a region near the center for these runs.
}
\end{table}

{\em Cocoon expansion:}
We begin with a study of the lateral extent of the cocoon, $r_c$. We find that the cocoon expands through the warm phase quite smoothly with time, as expected. Thus, we are able to study both the time dependence of $r_c$, and the dependence of the average cocoon expansion velocity ($v_c$) on the cloud parameters.

\begin{figure}
\begin{centering}
\hspace*{-0.7cm}\begin{tabular}{c}
{\mbox{\includegraphics[width=0.45\textwidth]{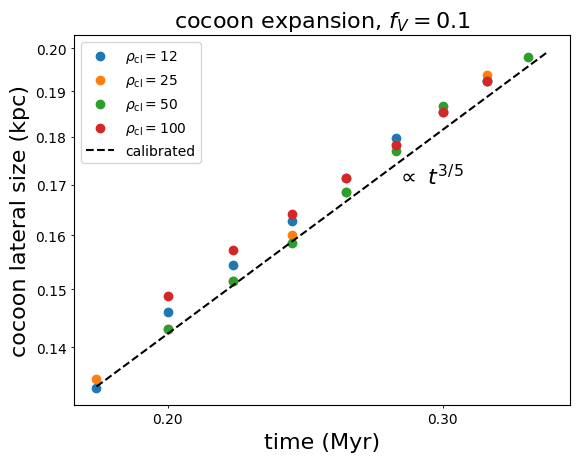}}}
\end{tabular}
\end{centering}
\caption{The time evolution of the cocoon size ($r_c$) for $f_V$ = 0.1. The cloud densities correspond to 12, 25, 50 and 100 $m_p$ cm$^{-3}$. These are in good agreement with the predicted scaling of $r_c \propto t^{3/5}$ and there is a very weak dependence on the cloud density. The dashed line shows the calibrated scaling for the cocoon radius from the left panel of Fig. \ref{fig:uniform_vs_t}.
} 
\label{fig:cocoon_01}
\end{figure}

\begin{figure}
\begin{centering}
\hspace*{-0.7cm}\begin{tabular}{c}
{\mbox{\includegraphics[width=0.45\textwidth]{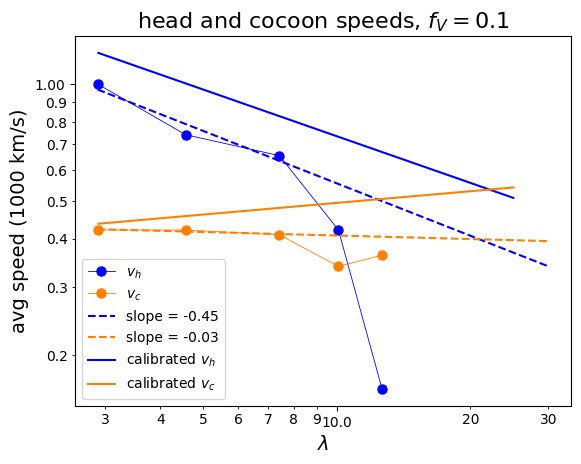}}}
\end{tabular}
\end{centering}
\caption{The markers show average head and cocoon velocities ($v_h$ and $v_c$) as a function of $\lambda$ for runs in Table \ref{table:pattern_const} with $f_V$ = 0.1. The cloud densities are $\rho_{\rm cl} = $ 12, 25, 50, 75 and 100 $m_p$ cm$^{-3}$. The solid lines show the speeds expected from our calibrated model for the jet head (blue; Eq. \ref{eq:vh_non-dissipated}) and cocoon (yellow; Eq. \ref{eq:vc_non-dissipated}), and the dashed lines are the power-law extrapolations of the simulation data. The head velocities are in good agreement with the predicted scaling of $v_h \propto \rho_{\rm eff}^{-2/5}$ in the non-dissipated regime (Eq. \ref{eq:vh_non-dissipated}).
} 
\label{fig:head_cocoon_speed_01}
\end{figure}

For the time dependence of $r_c$, our analytic model (section \ref{sec:analytic_estimates}) predicts that the instantaneous velocity in all regimes scales as $v \propto t^{-2/5}$. Integrating this, we obtain the prediction that $r_c \propto t^{3/5}$. From the plots of $r_c$ against time, shown in Fig. \ref{fig:cocoon_01}, we observe that the scaling is indeed close to the predicted value of 0.6. Moreover, the dependence of the cocoon velocity on cloud density (for a fixed density of the diffuse medium $\rho_H$) is very weak, consistent with Eq. \ref{eq:vc_non-dissipated}. The model predicts that in the non-dissipated regime, $v_c \propto \rho_{\rm eff}^{1/10}$ and in the dissipated regime, $v_c \propto \rho_{\rm eff}^{0}$ (since our simulations were performed keeping $\rho_H$ fixed, $\rho_{\rm eff}$ is proportional to $\lambda$). Such weak dependence is difficult to verify with simulations, however, we do confirm (from Fig. \ref{fig:head_cocoon_speed_01}) that the average $v_c$ stays almost the same (about 400 km s$^{-1}$ 
before the cocoon leaves the box)
over a large range of $\rho_{\rm eff}$ values (somewhat larger than values in Fig. \ref{fig:fV_vh_vc} because of an earlier time here for a smaller box). 

{\em Jet-head propagation:}
Since the dependence of cocoon expansion on the clumpiness of the medium (parameterized by $\lambda$) is quite weak, it is clear that the increase in jet dissipation with increasing $\lambda$ is largely due to the slowing down of the jet-head velocity, $v_h$. Since instantaneous $v_h$ in a clumpy medium can vary rapidly as the head interacts intermittently with clouds, it is difficult to compare its time evolution in simulations with analytic estimates. Hence, we focus on the time-averaged $v_h$ obtained from the simulations, in particular its dependence on $\rho_{\rm eff}$. 
The time-averaged head velocity is estimated as $v_h = (z_h(t_2) - z_h(t_1))/(t_2-t_1)$, where we consider $t_1$ to be the time right after the jet is fully collimated and $t_2$ to be the time right before the cocoon material leaves the simulation box. 

In the non-dissipated regime, the model predicts that the average head velocity should scale as $v_h \propto \rho_{\rm eff}^{-2/5}$. This is quite close to the results obtained from the simulations for low-density clouds, as shown in Fig. \ref{fig:head_cocoon_speed_01}. By extrapolating the scalings obtained in the non-dissipated regime, the critical value of $\lambda$ required for jet dissipation is found to be $\approx 20$ (see Fig. \ref{fig:head_cocoon_speed_01}) which is in good agreement with the calibrated prediction of 21.7 (Eq. \ref{eq:calibrated_lambda_crit}). However, we note that $v_h < v_c$ happens about a factor of 2 earlier than expected. We suspect that the reason for this shift is the lack of a hollowed-out region at the jet base for this set of simulations compared to the simulations used to calibrate $v_h$ and $v_c$ (section \ref{sec:uniform_test} and \ref{sec:rho_const}). This presence of dense clouds at the jet base delays jet collimation, invalidating the key assumption of our analytic models. However, we note that such effects become apparent only close to the dissipation regime. This indicates that our criterion for the jet dissipation, i.e. $\lambda_{\rm crit} \sim 20$ may require some revision for a realistic ISM. 

{While the simulations presented in this section show that the critical value of $\lambda$ may reduce (by a factor of 2) due to the presence of dense clouds in the immediate surroundings of the jet base, the distribution of dense ISM is uncertain for several reasons. Firstly, the local clouds may be evacuated by local supernovae activity or be arranged in a ring-like/disk-like fashion due to tidal interactions and angular momentum at the galactic centers \citep{Angles-alcazar2021}.} 
For our oversimplified clumpy ISM, with a larger simulation domain, we expect the influence of the delayed formation of collimation shock to reduce and an even closer match with the (calibrated) analytic prediction for jet dissipation. These runs also show the limitations of numerical simulations, because of finite resolution and small box-size, to capture all the relevant physics quantitatively. 

\section{Astrophysical implications}
\label{sec:astro_implications}

The primary motivation for this work was to apply the physics of self-collimated jets, well studied in the gamma-ray burst (GRB) literature, to galaxy-scale feedback. While AGN jet feedback simulations are commonplace, very often they lack sufficient resolution for producing a self-confined jet. Another differences from GRBs are the importance of clumpiness in the ISM and radiative cooling. In this paper, we generalize the jet-cocoon structure studied in \citet{bromberg11} to include the simplest prescription for a clumpy medium. The ISM disk is expected to be clumpy and there is a need for analytic scalings that can help apply the simulation results across a range of parameters in jet power, jet opening angle, ambient density, etc. In the following, we discuss the caveats and astrophysical implications of our study and comparison with related works.

\subsection{Jet dissipation criterion in a clumpy medium}
\label{sec:jet_diss}
For a given clumpy medium having cloud density $\rho_{\rm cl}$, volume filling fraction $f_V$ and cloud radius $R_{\rm cl} \approx 5\ {\rm pc}$, we describe here the procedure for computing the head and cocoon speeds according to our model. Firstly, $f_L$ is found using the reported Eq.  \ref{eq:f_Lf_V_relation} as $f_L = 0.6\ {f_V}^{1/3}$. Then, using Eq. \ref{eq:lambda_fL_rho} we get,
\begin{equation} \label{eq:lambda_fV_relation}
\lambda = \left( 0.6\ {f_V}^{1/3} \sqrt{\frac{\rho_{\rm cl}}{\rho_H}} + \left(1-0.6\ {f_V}^{1/3}\right)  \right)^2.
\end{equation}
This can then be used in equations such as Eqs. \ref{eq:vh_non-dissipated}, \ref{eq:vc_non-dissipated} to compute the time evolution of the jet-head and cocoon.

\begin{figure}
\begin{centering}
\hspace*{-0.7cm}\begin{tabular}{c}
{\mbox{\includegraphics[width=0.45\textwidth]{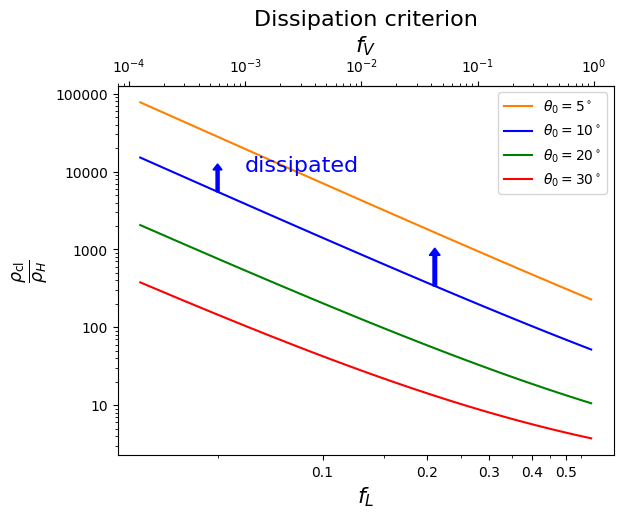}}}
\end{tabular}
\end{centering}
\caption{Curves showing the values of $f_L$ and $\rho_{\rm cl}/{\rho_H}$ above which the jet would be dissipated for four choices of the initial opening angle $\theta_0$, as predicted by our calibrated model, 
i.e., $\lambda_{\rm crit} = 21.7/\theta_{0,10}^2$ 
(Eq. \ref{eq:calibrated_lambda_crit}). The $f_V$ values (shown on top) corresponding to a given $f_L$ are obtained through the relation $f_L = 0.6\ {f_V}^{1/3}$.} 
\label{fig:dissipation}
\end{figure}

Using this relation, and the criterion for jet dissipation predicted by our calibrated model (Eq. \ref{eq:calibrated_lambda_crit}), we can compute the predicted critical values of $\rho_{\rm cl}/\rho_H$ and $f_L$ (or $f_V$), above which the jet will be dissipated. These threshold values are shown in Fig. \ref{fig:dissipation} for four choices of the initial opening angle of the jet, $\theta_0$. This figure suggests an easy way to evaluate if the jet is dissipated for given jet and ISM parameters.

\subsection{Excluded physics and implications}
\label{subsec:excluded-physics}
The main aim of our study is to obtain scaling relations for the structure of the cocoon blown by an AGN jet in a clumpy medium. In order to make progress analytically, we had to make extreme simplifications. Now we briefly mention the important physical effects that we excluded, and discuss their impact on jet-ISM+CGM interaction. 

\subsubsection{Radiative Cooling}

Section \ref{sec:uniform_analytic} shows that the cocoon pressure collimates the conical jet into a cylindrical geometry. However, radiative cooling at various locations within the cocoon (shocked ISM, shocked jet, mixed clump-cocoon gas) can reduce the cocoon pressure. Among other effects, a reduced cocoon pressure can decollimate the jet into a conical shape (see Eq. \ref{eq:Ht}). 

The cooling time (for free-free and line cooling) in terms of the pressure and temperature is given by 
\begin{equation}
    \label{eq:tcool}
    t_{\rm cool} \sim \frac{3}{2} \frac{k_B^2 T^2}{P \Lambda[T]} \sim 0.006~{\rm Myr}~ T_6^2 P_{-9}^{-1} \Lambda_{-22}^{-1},
\end{equation}
where $T_6=T[K]/10^6 $, $P_{-9} = P[{\rm dyn~cm^{-2}}]/10^{-9}$, $\Lambda_{-22} = \Lambda [{\rm erg~cm^3 s}^{-1}]/10^{-22}$. The solid lines in Fig. 7 of \citet{Kanjilal2021} show the variation of cooling time as a function of temperature in isobaric and isochoric conditions. The cooling time scales inversely with the pressure. In the absence of cooling, the density of the shocked homogeneous ISM is $4 \rho_H$ and the temperature of the outer cocoon is 
\begin{equation}
\label{eq:T_oc}
T_{\rm oc} = \frac{\mu m_p}{4 \rho_H} \frac{P_c}{k_B} = 1.2 \times 10^6~{\rm K} \frac{m_p {\rm cm}^{-3}}{\rho_H} \frac{P_c}{10^{-9} {\rm [dyn~cm^{-2}]}} .    
\end{equation}
The outer cocoon cooling time is longer at earlier times, $\propto P_c \Lambda^{-1} \propto t^{-4/5}$ (assuming a weak dependence of $\Lambda$ on $T$). The outer shock is expected to cool after $t>t_{\rm cool, oc}$, which for our fiducial parameters is $\approx 0.05$ Myr. By this time, the head moves beyond $\sim v_h t = 330$ pc (assuming a uniform ISM) into the lower density CGM where cooling is weaker. However, the cocoon still moves into the dense ISM. After this, most of the jet energy can be carried into the CGM which is not as clumpy and dense as the ISM. The passage of the jet into the lower density medium also leads to a sudden  lateral expansion of the jet+cocoon (e.g., \citealt{sutherland2007}).
We note that  while the dilute bubble contains most of the total energy injected by the jet, the radiative shell can still lose the majority of the injected energy for a dense ISM (e.g., see Fig. 8 of \citealt{Sharma2014} and Fig. 14 of \citealt{Yadav2017} for the importance of radiative losses in simulations of superbubble winds; see \citealt{Oey2009} for observational evidence).

Apart from the cooling of the shocked ISM, the mixed gas around cold clouds within the cocoon can also cool. As noted in section \ref{sec:jets_clumpy}, our clouds are isochoric early on and implode to become isobaric on a $\gtrsim 0.02$ Myr timescale (see middle panel of Fig. \ref{fig:fiducial}). %So 
Here, we assume the clouds to be roughly the same density as their original density in the ISM (see left panel of Fig. \ref{fig:fiducial}). The cloud-crushing time for the clouds interacting with the jet-head is much shorter than those embedded in the cocoon, since jet speed is much larger than the cocoon speed. The jet cloud density contrast is $\sim \rho_{\rm cl}/\rho_j \sim \rho_{\rm cl} \theta_0^2 v_j^2/(4 P_c) \sim (\theta_0^2/3) (\rho_{\rm cl}/\rho_H) (v_j/v_c)^2$ (see Eqs. \ref{eq:rho_j}, \ref{eq:P_c}), which is $\sim 5000$ at 1 Myr for our fiducial parameters. So the cloud-crushing time for clouds in the head is $\sim \sqrt{5000} R_{\rm cl}/v_j \sim 0.02$ Myr. For clouds interacting with the inner cocoon (rather than the jet), this time is longer since the relative speed is smaller. The mixed gas (with a higher temperature) can be considered to have a temperature $\sim \sqrt{T_{\rm ic} T_{\rm cl}}$ (\citealt{Begelman1990}), where the inner cocoon temperature $T_{\rm ic} \sim \mu m_p P_c/(k_B \rho_{\rm ic}) \sim \mu m_p v_j^2 \theta_0^2 / (4 k_B) \sim 10^8$ K (the inner cocoon density $\rho_{\rm ic}$ is assumed to be $\sim \rho_j$). Thus, the mixed gas is at $\sim 10^5$ K and the cooling time of this gas at the cocoon pressure is $\sim 6$ yr (plugging $T_6 = 0.1$ and $\Lambda_{-22} = 10$ in Eq. \ref{eq:tcool})! Since the cooling time of mixed gas is very short, loss of cocoon pressure happens on the mixing time $\sim t_{\rm cc} \sim (R_{\rm cl}/v_c) (\rho_{\rm cl}/\rho_c)^{1/2} \gtrsim 0.05$ Myr (mixed gas in the inner cocoon has even longer cooling time) rather than the cooling time of the mixed gas. Thus, both the cooling of the shocked ISM and that of the mixed gas, and subsequent loss of cocoon pressure, start to happen after 0.05 Myr. Although the volume occupied by mixed gas is small, it may still radiate away most of the cocoon thermal energy because of high radiative efficiency. 
Note that our estimates are very crude and must be tested with jet-ICM simulations with radiative cooling. 
Despite radiative cooling losses, our analytic framework can still be used by replacing the jet luminosity $L_j$ in analytic estimates by $\eta L_j$, where $\eta < 1$ is an efficiency factor that accounts for radiative losses.

If the jet is able to drill through the dense ISM by this time, it is not expected to choke due to catastrophic radiative losses. The Compton cooling time for the cocoon with CMB and stellar light as seed photons is much longer (we assume the central AGN to have a luminosity much smaller than the Eddington limit, as appropriate for most nearby AGN). The electron-proton energy equipartition time and the radiative recombination time for the dense/mixed gas are short, so that the equilibrium cooling function can be used. It will be most interesting to extend our simulations to include cooling and realistic ISM+CGM gas distribution.

\subsubsection{Disk-CGM transition: Comparison with J1316+1753}
The other unrealistic simplification that we have made for ease of analytic treatment is that of statistical homogeneity and isotropy. In reality, the warm/cold ($10^4/100$ K) ISM is strongly stratified with the disk scale height much smaller than the radial scale length. In a stratified ISM, once the cocoon expands beyond a disk scale height, it accelerates vertically and slows its expansion into the disk. After this, the cocoon energy is preferentially deposited in the less dense/non-radiative CGM and our assumption of a uniform medium breaks down even qualitatively. Our scalings do tell us when we expect the cocoon to break out of the ISM disk.

Numerical simulations (\citealt{Mukherjee2018,Tanner2022}) show that even jets that are launched into the clumpy ISM disk at a large inclination angle create a  bipolar outflow symmetric relative to the galactic center. Once cocoon reaches close to the disk scale height, it takes the path of lowest column along the galaxy minor axis. This is also consistent with multiwavelength observations (e.g., \citealt{Venturi2021, Girdhar2022}) which show the warm nebular gas with large velocity dispersion perpendicular to the ISM disk.

{\em Applying our model to J1316+1753 jet:} Our formalism of the jet encountering a larger effective density compared to the cocoon (section \ref{sec:jet_clumpy_analytic}) can also be applied to a jet propagating into the ISM disk and the cocoon expanding into the lower density warm/hot medium. Assuming that the jet is launched into the dense ISM disk, $\rho_{\rm eff} \approx \rho_{\rm mid} \sim 3 \times 10^4 m_p$ g cm$^{-3}$ (the density in the ISM mid-plane) and $\rho_H \approx \rho_{\rm diffuse} \sim 100 m_p$ g cm$^{-3}$, where we quote the density estimates in nearby  quasar J1316+1753 from \citet{Girdhar2022}. The electron density of nebular phase is estimated from [SII] doublet to be $\sim {\rm few}~100$ cm$^{-3}$ (we assume a lower value because the ambient medium density is $4\times$ smaller than the shocked density in a non-radiative shock) and molecular phase (at few 10s of K) in the midplane is assumed to be in pressure equilibrium with the warm phase. For J1316+1753 the jet-head is clearly well within the cocoon which expands along the disk minor axis. For the anisotropic medium (within disk plane versus perpendicular to it) we consider here, the estimates in section \ref{sec:analytic_nondiss} are valid even for dissipated jets like J1316+1753 because the cocoon is preferentially moving into the low-density CGM and the jet-head is obstructed by molecular gas in the disk. The length factor is expected to be $f_L \approx 1$ for jet propagating into the molecular disk and thus $\lambda \equiv \rho_{\rm eff}/\rho_H \sim \rho_{\rm mid}/\rho_{\rm diffuse} \sim 300$. Also, $f_V \ll 1$ since most of the cocoon volume is occupied by the diffuse warm phase away from the midplane. Assuming $L_j \approx 10^{44}$ erg s$^{-1}$ (estimated from radio luminosity in \citealt{Girdhar2022}), the fiducial $\theta_0 \approx 10^\circ$, $\rho_H \sim 100 m_p$ cm$^{-3}$, and the measured cocoon radius $r_c \sim 7$ kpc, we can calculate the age of this cocoon from integrating Eq. \ref{eq:vc_non-dissipated} and using the calibration from Fig. \ref{fig:uniform_vs_t} as $\sim 75$ Myr (where we have plugged in $\lambda=\rho_{\rm mid}/\rho_{\rm diffuse}= 300$ in Eq. \ref{eq:vc_non-dissipated}). The numerically calibrated $r_c/z_h \approx 0.215 \theta_{0,10} \lambda^{1/2}$ from Eq. \ref{eq:diss_ratio_app} with the measured $r_c/z_h \approx 7/1.5$ gives $\lambda \sim 470$. This inferred $\lambda$ is close to the ratio $\rho_{\rm mid}/\rho_{\rm diffuse} \sim 300$. Thus, our analytic model is self-consistent and has the potential to be applied quantitatively to realistic observations of AGN jet driven cocoons.

The other noteworthy feature of J1316+1753 is that the cocoon (traced by [OIII] emission) is symmetric relative to the galactic center rather than around the jet hot spots offset by $\approx$ 1.5 kpc from the center. This suggests that the galaxy's minor axis provides the path of least resistance (with lowest column density) for the cocoon to move. A low density CGM along the galaxy minor axis is indeed seen in simulations (e.g., \citealt{Mukherjee2018, Yang2024, Nelson2019})\footnote{In these works, AGN feedback is injected in very different ways, but the propagation of cocoon preferentially along the galaxy minor axis is a generic outcome.} and inferred from observations (\citealt{Martin-Navarro2021, Zhang2022}). This implies that Fermi/e-Rosita bubbles in the center of Milky Way, which display left-right and top-down symmetry, may be produced by inclined jets that dissipate within $\lesssim 1$ kpc from the center as in J1316+1753 (see \citealt{Sarkar2023} and section \ref{sec:FEBs} for further discussion).  

\subsubsection{Realistic clumpy medium}
To minimise the number of parameters needed to describe the clumpy medium, we simply assume the ISM to have uniformly randomly distributed spherical clouds with a specified density contrast, volume filling fraction, and cloud size. However, the ideas presented for our simple cloudy ISM are generically applicable. Namely, the jet interacts strongly with the clouds on its path and the clouds away from the jet-beam are slowly engulfed by the hot cocoon.

A more realistic distribution is a multiphase ISM+CGM with a distribution across a range of densities. If the volume-PDF over density has separate peaks (e.g., $t=0$ PDF in Fig. 2 of \citealt{mukherjee2016}, and even at later times) then we can still calculate the effective density encountered by the jet as given by Eq. \ref{eq:lambda_fL_rho} with $\rho_H$ as the density peak value of the phase with largest $f_V$, $\rho_{\rm cl}$ as the density peak value of the dense phase, $f_L \approx 0.6 f_V^{1/3}$ (Eq. \ref{eq:f_Lf_V_relation}), where $f_V$ is the volume fraction of the dense phase. For \citealt{mukherjee2016} then,  $\rho_H \approx 0.1$ cm$^{-3}$ and $\rho_{\rm cl} \approx 100$ cm$^{-3}$, and $f_V \approx 0.045$ (mentioned in [iii] of their section 3) gives $f_L \approx 0.21$). Thus, our Eq. \ref{eq:lambda_fL_rho} gives $\lambda \approx 55$, which according to Eq. \ref{eq:calibrated_lambda_crit} is well within the dissipated regime. Indeed, the jet-head in the second row of Fig. 5 (time = 0.56 Myr) of \citet{mukherjee2016} is dissipated within a spherical cocoon. Later, when the jet-head exits the ISM and enters the non-clumpy CGM, the cocoon becomes anisotropic, and the jet is not dissipated.

While a multi-peaked volume PDF appears to be a decent approximation for a two-phase medium that we consider, the volume-PDF can be more general, but our idea that the jet encounters a higher density than the cocoon still applies. In this case, the effective density parameter $\lambda$ may be given by
\begin{equation}
    \lambda 
    \sim \frac{\langle \sqrt{\rho} \rangle_L^2}{\rho_H},
\end{equation}
where $\langle \rangle_L$ represents the average over the length-PDF (related to the volume-PDF; recall from section \ref{sec:uniform_clouds_analytic} that the length covered by clouds rather than their volume fraction matters for effective density encountered by the jet) and $\rho_H$ is the density corresponding to the global maximum of the volume-PDF. 

\subsubsection{Magnetic fields}
We model the jets hydrodynamically, but the jets are expected to be magnetically dominated. A large enough magnetic field dominated by the azimuthal component can become unstable to the kink instability (e.g., \citealt{tchekhovskoy2016, Mukherjee2020}), which can dissipate the jet internally rather than due to interaction with the ambient medium. It will be useful to quantify the relative role of magnetic instabilities and clumpiness in dissipating the AGN jet, but this investigation is left for future.

\subsection{Jet-ISM interaction time \& AGN feedback}

As long as the jet is active, the jet-head can drill through a dense cloudy medium. However, after the jet is off (called choked jet in \citealt{Sarkar2023}; see also \citealt{Yang2022}), the inner cocoon is crushed by the shocked ISM and shock+bubble evolution is governed by ambient density gradient and buoyancy. For a sufficiently energetic jet event, most of the energy is deposited in the CGM (and not the ISM) via shocks and mixing. If the jet switches off before the cocoon breaks out into the CGM, most of the feedback energy is deposited in the dense ISM (absent in massive elliptical galaxies) and may be radiated away without much long-term self-regulation of star formation. In this case, localized positive feedback is likely. A few eddy turnover times after the jet switches off, the molecular gas can cool and form stars prodigiously, and more molecular gas can condense out from colliding warm clouds (e.g., \citealt{Heitsch2006}). Thus, the CGM seems to  where jet energy must be deposited for negative feedback to dominate on average (e.g., see Fig. 6 in \citealt{Fielding2018}, which shows an increase in mechanical efficiency once the superbubble breaks out of the ISM into the CGM; see also Fig. 3 in \citealt{Shchekinov2018}).

Let the effective path length that the jet-head traverses within the ISM be $l_{\rm ISM}$, and the time for which the jet is confined inside the ISM of the host galaxy (or the jet-ISM interaction time) be denoted by $T_{\rm ISM}$. Then
\begin{equation}
l_{\rm ISM} = \int v_h \,dt
\end{equation}
or,
\begin{equation}
\begin{split}
l_{\rm ISM} &= \frac{10}{3 (4 \pi)^{1/5}} (1-f_V)^{-1/5}\\ 
&\times (\theta_0 \sqrt{\lambda})^{-4/5} \left(\frac{L_j}{\rho_H} \right)^{1/5} {T_{\rm ISM}}^{3/5},
\end{split}
\end{equation}
assuming the jet to be in the non-dissipated regime (see section \ref{sec:analytic_nondiss}; Eq. \ref{eq:vh_non-dissipated}). Now assuming a constant disk thickness $2H$ and the inclination angle of the jet with respect to the disk normal to be $\theta_j$, we obtain \(l_{\rm ISM} = H/ \cos \theta_j\). Thus, our model gives the following prediction for this ISM interaction time 
\begin{equation}
\begin{split}
T_{\rm ISM} &= \left(\frac{3}{10}\right)^{5/3} (4 \pi (1-f_V))^{1/3}\\ 
&\times \left( \theta_0 \sqrt{\lambda} \right)^{4/3} \left(\frac{\rho_H}{L_j} \right)^{1/3} \left( \frac{H}{\cos \theta_j} \right)^{5/3}.
\end{split}
\end{equation}
The physical significance of $T_{\rm ISM}$ is that it sets the typical time that is available for the jet to interact directly with the host ISM, before it drills through this clumpy ISM. This is thus the time for which direct jet feedback can operate (e.g., by driving turbulence in the ISM which can affect star formation, as studied by \citealt{mukherjee2016}), and also the time for which the jet and cocoon structure can be affected by the properties of the host ISM. At later times after the jet drills through the ISM, most of the energy is deposited outside the galactic disk, and then one can expect that the jet feedback would be mostly in the maintenance mode, where the jet energy prevents cooling flows onto the galaxy (\citealt{fabian12}). Substituting typical values for the relevant quantities, we obtain
\begin{equation}
\begin{split}
\label{eq:TISM}
T_{\rm ISM} &= 
0.026\ {\rm Myr}\ (1-f_V)^{1/3} \\
&\times \left(\frac{\theta_{0,10}^4}{L_{j,42}} \right)^{1/3} \left( \rho_{H,m_p} \lambda^2 \right)^{1/3} \left( \frac{H_{200}}{\cos \theta_j} \right)^{5/3},
\end{split}
\end{equation}
where $H_{200}$ is the height of the clumpy medium, in units of 200 ${\rm pc}$. Here we have used the numerical calibration of $z_h$ from Fig. \ref{fig:uniform_vs_t}.

This shows, for instance, that nearly dissipated jets facing $\lambda \approx 100$ can be confined within the ISM for a time $\gtrsim 5\ {\rm Myr}$, provided they are low-power ($L_j \lesssim 10^{41}\ {\rm erg\ s^{-1}}$) jets launched with a sufficiently large ($\gtrsim 65^{\circ}$) angle of inclination with respect to the disk normal This time further increases if jets with lower power or wider opening angles are considered. 
For jet-ISM interaction, the jet duty cycle (fraction of time that the jet is ON) is important but even more important is the duration of individual jet events. Longer ON states of the radio jet makes it easier for jet to deposit its energy in the CGM rather than being stopped in the ISM.

\subsection{Anisotropy of jet power \& AGN feedback in simulations}
\label{sec:jets_cosmo}

For low-power, inclined jets, if the time for which the jet is active is shorter than or comparable to $T_{\rm ISM}$, then at large scales, the outflow is mainly due to the expansion of the overpressured cocoon into the galactic halo, which thus takes the form of a relatively isotropic bubble, whose energy is mostly in the form of thermal energy close to the disk. On the other hand, if the jet lifetime is much longer than $T_{\rm ISM}$, then the jet-head can break out of the ISM with ease, following which it moves much faster in the diffuse CGM than the lateral expansion of the cocoon (as the head is no longer obstructed by dense clumps). This results in the jet feedback being highly anisotropic, and the energy flux being dominated by the kinetic energy of the collimated jet that is dissipated in a hot spot at the edge of an anisotropic cocoon, as in FRII jets. These considerations motivate subgrid models for mechanical feedback by AGN jets in cosmological galaxy formation simulations.

The spatial resolution in cosmological simulations is $\gtrsim 100$ pc (e.g., see Table 2 of \citealt{Vogelsberger2020}), much larger than the jet radius for typical parameters (Eq. \ref{eq:Rj_uniform_pred}). Thus, such simulations cannot evolve self-consistent jets confined by the cocoon pressure. The AGN jet implementations in cosmological simulations should be considered as effective models that inject enough energy in the CGM to prevent runaway cooling flows. However, the cocoon structure and microphysics of energy dissipation are unresolved.

Since the ratio of cocoon radius and the jet-head is $\sim  \theta_0 \ll 1$ (see Eq. \ref{eq:diss_ratio_app}), the energy injection due to AGN jets should be anisotropic with the energy propagating primarily in the jet direction. This is accomplished by injecting AGN jets with a specified mass, momentum, and energy flux in the radial direction but confined to a narrow angle $\sim \theta_0$ (e.g., as done in \citet{Prasad2015, Li2015, Yang2016}). In fact, a range of feedback prescriptions, as long as they deposit energy anisotropically (or become effective anisotropic because of the ISM; e.g., \citealt{Sijacki2007,Nelson2019}), are able to produce galaxy and CGM properties in broad agreement with observations. If AGN energy is deposited isotropically, the jets do not propagate far from the cluster core but are smothered by excessive cooling in the dense cluster core with a short cooling time (\citealt{Meece2017}). For anisotropic energy deposition by jets, most of the jet energy is deposited at large radii, which prevents excessive cooling over cosmological timescales. Only anisotropic jets allow energy to be transported to large radii and control long term cooling flows, even though cooling episodes lead to short starbursts. 

Our analytic framework provides a framework for injecting isotropic or jetted AGN feedback depending on the clumpiness assumed at scales below the numerical resolution. Eqs. \ref{eq:f_L_def}, \ref{eq:lambda_fL_rho}, \ref{eq:calibrated_lambda_crit} give the threshold for clumpiness  ($\lambda$) beyond which the cocoon is expected to be isotropic, and a physically consistent feedback should be isotropic. For smaller clumpiness, energy and mass can be injected over a narrow angle $\sim \theta_0$, corresponding to a cocoon launched by a narrow jet in a uniform medium. We can make the subgrid energy injection prescriptions more sophisticated by
smoothly interpolating the anisotropy of injection of AGN power between the uniform and the fully dissipated regime, depending on the clumpiness of the ISM as quantified by $\lambda$.

\subsection{Fermi/eROSITA bubbles}
\label{sec:FEBs}
The origin of the Fermi/eROSITA bubbles (FEBs) has been discussed in the context of supernovae driven wind or from an AGN activity from \sgra \citep{Su2010, Zubovas2012, Crocker2015, Mou2015, Sarkar2015b, Sarkar2017, Sarkar2019, Zhang2022, Mondal2022, Yang2022}. While the supernovae-driven wind naturally follows the density gradient of the galaxy and produces symmetric FEBs, the AGN jet-driven events are not expected to follow the density gradient unless the jet is dissipated within the ISM. It was argued by \cite{Sarkar2023} that any past jet activity at the \sgra was most probably inclined with respect to the Galaxy rotation axis and that any such inclined jets would produce asymmetric FEBs, unlike the observed ones. A successful jet-driven FEB model, therefore, requires the jet to be dissipated. 

The current paper provides a framework to check if any past jets in the MW could have been dissipated. The MW ISM is characterized by a typical volume filling density of $\rho_H \sim 1$ \mpcc in which more dense gas such as diffuse H$_2$ phase ($\rho_{\rm cl} \sim 100$ \mpcc, $f_V \sim 10^{-3}$) or dense H$_2$ phase ($\rho_{\rm cl} \sim 10^4$ m$_p$ cm$^{-3}$, $f_V \sim 10^{-4}$) reside \citep{Draine2011}. Now, combining Eq.  \ref{eq:lambda_fL_rho} and that $f_L \approx 0.6 f_V^{1/3}$ (from Fig. \ref{fig:fL_vs_fV}), we estimate that the MW ISM can be described by $\lambda \sim 2.4$ and $\lambda \sim 14$ for the diffuse H$_2$ and dense H$_2$ phases, respectively. A comparison to Eq.  \ref{eq:calibrated_lambda_crit} shows that in MW ISM, $\lambda < \lambda_{\rm crit}$. This means that the MW ISM is not expected to strongly dissipate any jets from \sgra and, jets should blow bubbles that are asymmetric relative to Milky Way's minor axis.

One possibility, however, remains. It is the interaction of the jet with the central molecular zone (CMZ). The CMZ is an elliptical ring (semi-major axis $\approx 100$ pc) of dense molecular gas around the Galactic Center \citep{Molinari2011}. The dense gas in the CMZ has a density of $\sim 10^3$ \mpcc \citep{Morris1996, Molinari2011, Henshaw2016} and lies on the Galactic plane. Given that the `best-bet' rotation axis of the central SMBH lies close to the disc \citep{EHT2022}, it could well be possible that the past jet had interacted with the CMZ. Now, considering that the CMZ ring has an inner and outer radius of $60$ pc and $100$ pc and that its height is $\sim 20$ pc, the volume filling factor for the CMZ within the ISM scale height is $\sim 0.06$ (assuming ISM scale height of $200$ pc). This implies that the CMZ has $\lambda \sim 70 \gg \lambda_{\rm crit}$. Therefore,  it is plausible that if the MW jet interacts with the CMZ, the jet could be dissipated to produce the symmetric FEBs. Such an interaction would, however, be imprinted on the structure of the CMZ as breaks in dense gaseous rings. A detailed study of such a scenario remains for future work. Yet another possibility to produce symmetric FEBs is that of choked jets discussed in \citet{Sarkar2023}.
However, such a choked jet would produce a stronger shock compared to the observations and is ruled out as a possible explanation for the FEBs.

\subsection{Ubiquity of low power jets in massive galaxies}
The earliest detected radio jets and bubbles were naturally very powerful, but the most common radio jets appear to be compact and lower power, not only in elliptical galaxies and clusters but also in disk galaxies (e.g., \citealt{baldi18, Sabater2019}). Moreover, luminous radio jets ($L_{\rm 150 MHz} \gtrsim 10^{23}$ W Hz$^{-1}$, or $L_{\rm jet} \gtrsim 10^{43}$ erg s$^{-1}$) are exceedingly rare in star-forming galaxies but are almost exclusively found in massive elliptical galaxies (see Fig. 4 in \citealt{Sabater2019}). Most importantly, galaxies with stellar mass $\gtrsim 10^{11} M_\odot$ almost always have $L_{\rm 150 MHz} \gtrsim 10^{21}$ W Hz$^{-1}$ or $L_{\rm jet} \gtrsim 2 \times 10^{41}$ erg s$^{-1}$. A larger jet power in more massive galaxies/halos implies a tight feedback coupling between the cooling of the CGM/ICM and accretion power in massive halos.

Recent observations have shown that most of the radio sources in the local universe are compact jets with linear sizes $\lesssim 5\ {\rm kpc}$, in contrast to the extended radio emission from FR I and FR II type jets. These have been termed as FR 0 jets (\citealt{baldi18}). As discussed in \cite{baldi18}, these sources have been observed to be much more numerous than that expected if these are simply young FR I jets. On the other hand, their galactic environment seems to be similar to that of FR I jets. This suggests that they are intrinsically different, possibly having lower power or shorter active lifetimes compared to FR I jets. These observations imply that such FR 0 jets are likely to spend a significant fraction of the active lifetime within the host galaxy. As a result, the dynamics of their jet and cocoon, and the resulting feedback on the galaxy, would be significantly affected by the interaction of the jet with the ISM. 

Elliptical galaxies typically lack a dense ISM and the jets propagate in the hot ICM/CGM. The CGM pressure/temperature scales with the halo mass as $\propto M_{\rm halo}^{2/3}$, but the CGM density within $\lesssim 10$  kpc is similar (e.g., see Fig. 2 in \citealt{Sharma2012}). A cocoon propagating in the hot CGM will eventually reach pressure equilibrium with the CGM and then evolve buoyantly. Using Eqs. \ref{eq:vh_uniform_pred}-\ref{eq:P_uniform_pred}, the cocoon radius when it reaches the ambient pressure $P_{a}$ is,
\begin{equation}
    R_{c, \rm buoy} \approx 1.6~{\rm kpc} \left( \frac{\theta_{0,10}^2 L_{j,42}^2 \rho_{H,m_p}}{P_{a,-10}^3} \right)^{1/4},
\end{equation}
where the ambient pressure is normalized to $10^{-10}$ dyn cm$^{-2}$, comparable to the ICM temperature of $10^7$ K and number density of 0.1 cm$^{-3}$.\footnote{Plugging in the typical CGM parameters for Milky Way, $\rho_{H,m_p} \sim 0.01$ and $P_{a,-10} \sim 0.01$, $R_{c, \rm buoy} \sim 16$ kpc, of order the size of Fermi/eRosita bubbles.} Note that this estimate is only weakly dependent on the density profile for any reasonable model of the ICM (see Appendix B.1 of \citealt{bromberg11}). The corresponding head location is $z_h \approx 7.4$ kpc (using Eq. \ref{eq:diss_ratio_app}). This implies that the mechanical power of FR 0 jets with size $\lesssim 5$ kpc is $\lesssim 10^{42}$ erg s$^{-1}$. This limit corresponds to $L_{\rm 1.5 GHz} \approx 2.5 \times 10^{21}$ W Hz$^{-1}$ (see equation relating cavity power and radio power in \citealt{Sabater2019}). Reading off from Fig. 5 of \citet{Sabater2019}, $\gtrsim 85$\% of galaxies with $M_\star \lesssim 10^{11} M_\odot$ have radio luminosity larger than this value, thus making FR 0s the most numerous radio jets in the local Universe, quantitatively consistent with \citet{baldi18}. 

While the number of galaxies is dominated by the lowest radio power jets, the total jet power is dominated by the galaxies at the cut-off jet radio power, which for star-forming galaxies is $L_{\rm 1.5 GHz} \sim 10^{22}$ W Hz$^{-1}$ and for AGN sample is $\sim 10^{25}$ W Hz$^{-1}$ (e.g., see Fig. 4 in \citealt{Sabater2019}) corresponding to $P_j \sim 10^{42},~10^{44}$ erg s$^{-1}$, respectively.

\section{Summary}
\label{sec:summary}
Here is a very brief summary of our paper:

\begin{enumerate}
    \item We extend the analytic model of a self-collimated hydrodynamical conical jet injected into a uniform medium to a clumpy medium characterized by the density ratio of the cold clouds relative to the diffuse medium $\rho_{\rm cl}/\rho_H$, the volume filling fraction $f_V$, and the cloud radius $R_{\rm cl}$. We calibrate the cocoon radius and the jet-head height with numerical simulations (see Fig. \ref{fig:uniform_vs_t}). The ratio of the cocoon radius to the jet-head location depends only upon the jet injection angle $\theta_0$ (Eq. \ref{eq:diss_ratio_app}). The key physical insight behind our model is that the supersonic jet encounters a much higher effective density (given by Eq. \ref{eq:lambda_fL_rho}) compared to the subsonic cocoon that effectively encounters the diffuse medium. The ratio of the effective density encountered by the jet and the cocoon $\lambda$ depends on the density contrast between clouds and the diffuse medium ($\rho_{\rm cl}/\rho_H$) and the length fraction in clouds along the jet-beam $f_L$. Our numerical calibration shows that $f_L \approx 0.6 f_V^{1/3}$ for the range of cloud sizes that we have explored. Using the analytic model and calibration with numerical simulations, we find the criterion for strong jet dissipation (Eq. \ref{eq:calibrated_lambda_crit}; see also section \ref{sec:jet_diss}) as a function of $\lambda$ and $\theta_0$. Jet dissipation corresponds to the jet being stopped by the clouds along the jet-beam well inside the cocoon.
    \item Numerical simulations suggest two qualitatively different regimes for the interaction of an AGN jet with the ambient medium: (i) anisotropic regime in which the jet-head moves farther than the cocoon and the cocoon is elongated along the jet; and (ii) dissipated regime in which the head velocity is smaller than that of the cocoon and the cocoon is isotropic. The dissipated jet regime occurs for sufficiently dense clumps and a large enough volume filling fraction, the threshold being quantitatively given by $\lambda \gtrsim 20$ where $\lambda$ is given by Eq. \ref{eq:lambda_fV_relation}. These two regimes (see section \ref{sec:jets_cosmo}) suggest a simple subgrid model (anisotropic versus isotropic energy injection) for the implementation of mechanical AGN feedback in galaxy formation simulations. We apply our calibrated analytic models to a host of astrophysical systems from Fermi/eRosita bubbles to Seyfert/quasar and FR 0 jets in section \ref{sec:astro_implications}.
\end{enumerate}

%\ks{
Overall, we present a theoretical framework, for the first time, to describe the dissipation of AGN jets in a clumpy medium and discuss its implications for different modes of jet feedback in galaxies and galaxy clusters. %}
Our models/simulations are still missing some of the key physical processes such as radiative cooling, magnetic fields, and realistic gravitational and density fields. Strong azimuthal magnetic fields within the jet can make it unstable to magnetic instabilities such as the kink and tearing modes. Similarly, radiative cooling may take away most of the jet energy and the cocoon expansion can be slowed. We aim to study these important effects in the future. 

\section*{Acknowledgements}
We thank Dipanjan Mukherjee for helpful discussions. We acknowledge National Supercomputing Mission (NSM) for providing computing resources of Param Pravega at IISc. KCS is partially supported by the German Science Foundation via DFG/DIP grant STE/ 1869-2 GE/ 625 17-1 in Israel. JMS acknowledges support from the Eric and Wendy Schmidt Fund for Strategic Innovation.

\bibliography{references.bib}
\bibliographystyle{aasjournal}

\end{document}